\documentclass{aa}
\usepackage{natbib}
\usepackage[utf8]{inputenc}
\usepackage[T1]{fontenc}
\usepackage{amsmath}
\usepackage{amssymb}
\usepackage{geometry}
\usepackage{graphicx}
\usepackage{bm}
\usepackage{indentfirst}
\usepackage{hyperref}
\usepackage{caption}
\usepackage{subfig}
\usepackage{xcolor}
\usepackage{txfonts}
\usepackage{ulem}

\bibpunct{(}{)}{;}{a}{}{,}

\newcommand*\diff{\mathop{}\!\mathrm{d}}
\newcommand{\vv}{\mathrm{v}}

\begin{document}

\title{Modelling the asymmetries of the Sun's radial $p$-mode line profiles}

\author{J. Philidet\inst{\ref{inst1}} \and K. Belkacem\inst{\ref{inst1}} \and R. Samadi\inst{\ref{inst1}} \and C. Barban\inst{\ref{inst1}} \and H.-G. Ludwig\inst{\ref{inst2},\ref{inst3}}}

\institute{LESIA, Observatoire de Paris, PSL Research University, CNRS, Universit\'e Pierre et Marie Curie, Universit\'e Paris Diderot, 92195 Meudon, France \label{inst1} \and Zentrum für Astronomie der Universität Heidelberg, Landessternwarte, Königstuhl 12, 69117 Heidelberg, Germany\label{inst2} \and GEPI, Observatoire de Paris, PSL Research University, CNRS, Universit\'e Pierre et Marie Curie, Universit\'e Paris Diderot, 92195 Meudon, France\label{inst3}}

\abstract
{The advent of space-borne missions has substantially increased the number and quality of the measured power spectrum of solar-like oscillators. It now allows for the $p$-mode line profiles to be resolved and facilitates an estimation of their asymmetry. The fact that this asymmetry can be measured for a variety of stars other than the Sun calls for a revisiting of  acoustic mode asymmetry modelling. This asymmetry has been shown to be related to a highly localised source of stochastic driving in layers just beneath the surface. However, existing models assume a very simplified, point-like source of excitation. Furthermore, mode asymmetry could also be impacted by a correlation between the acoustic noise and the oscillating mode. Prior studies have modelled this impact, but only in a parametrised fashion, which deprives them of their predictive power.}
{In this paper, we aim to develop a predictive model for solar radial $p$-mode line profiles in the velocity spectrum. Unlike the approach favoured by prior studies, this model is not described by free parameters and we do not use fitting procedures to match the observations. Instead, we use an analytical turbulence model coupled with constraints extracted from a 3D hydrodynamic simulation of the solar atmosphere. We then compare the resulting asymmetries with their observationally derived counterpart. }
{We model the velocity power spectral density by convolving a realistic stochastic source term with the Green's function associated with the radial homogeneous wave equation. We compute the Green's function by numerically integrating the wave equation and we use theoretical considerations to model the source term. We reconstruct the velocity power spectral density and extract the line profile of radial $p$-modes as well as their asymmetry.}
{We find that stochastic excitation localised beneath the mode upper turning point generates negative asymmetry for $\nu < \nu_\text{max}$ and positive asymmetry for $\nu > \nu_\text{max}$. On the other hand, stochastic excitation localised above this limit generates negative asymmetry throughout the $p$-mode spectrum. As a result of the spatial extent of the source of excitation, both cases play a role in the total observed asymmetries. By taking this spatial extent into account and using a realistic description of the spectrum of turbulent kinetic energy, both a qualitative and quantitative agreement can be found with solar observations performed by the \textit{GONG} network. We also find that the impact of the correlation between acoustic noise and oscillation is negligible for mode asymmetry in the velocity spectrum.}
{}


\maketitle

\section{Introduction}\label{sec:intro}

Solar-like oscillations are known to be stochastically excited and damped by turbulence occurring close to the surface of low-mass stars (see e.g. \citealp{GK77a, GK77b} or \citealp{samadiBS15} for a review). The power spectral density of such oscillations is expected to feature a Lorentzian-shaped peak centred around their eigenfrequencies. This idealised line profile has been extensively used to fit observations \citep[see e.g.][]{jefferies91}. However, as the resolution reached in helioseismic measurements (both ground-based and space-borne) has increased, it has become apparent that the observed line profiles feature a certain degree of asymmetry (see e.g. \citealp{duvall93} for observations made at the geographic South Pole; \citealp{toutain98} for data from the \textit{MDI} and \textit{SPM} instruments aboard the \textit{SOHO} spacecraft).

Since the discovery of this skew symmetry in solar $p$-mode line profiles, several studies have been devoted to explaining this feature. In particular, it had been recognised early on that a source of excitation that is highly localised compared to the mode wavelength (which we refer in the rest of the paper as `source localisation') could lead to a certain degree of mode asymmetry, depending on the position of the source \citep{gabriel92, duvall93, gabriel93, abramsK96, roxburghV95, roxburghV97}. Line profile asymmetries have then been used to infer some properties of the turbulent source, especially its radial location and its multipolar nature \citep[see  e.g.][]{roxburghV97, nigamCORREL98}.

Furthermore, \citet{duvall93} noticed an inversion of the sense of asymmetry between spectrometric and photometric measurements, with line profiles in the velocity spectrum featuring more power in their low-frequency wing than in their high-frequency wing and vice-versa for line profiles in the intensity spectrum. Since intensity perturbations were expected to be proportional to velocity perturbations, one would have expected the asymmetries to be the same. Many hypotheses were posited to explain this puzzling result. \citet{duvall93} suggested that it was due to non-adiabatic effects lifting the proportionality relationship between the two kinds of perturbations (fluid displacement and temperature) but this hypothesis was later contradicted by \citet{rastB98}. Non-adiabaticity was brought up again later on by \citet{georgobiani03} who suggested that the explanation resided in radiative transfer between the mode and the medium. Indeed, the observed radiation temperature corresponds to the gas temperature at local optical depth $\tau = 1$. But optical depth depends on opacity, which non-linearly depends on temperature. Therefore, the temperature fluctuations due to the oscillating mode entails opacity fluctuations, which in turn impacts the  `observed'\ radiation temperature. Given the non-linear nature of the $\kappa - T$ relation, this modulation decreases the observed temperature fluctuations more significantly in the low-frequency wing of the mode than in its high-frequency wing. Since this radiative transfer does not impact the velocity measurements, this could explain the asymmetry reversal between velocity and intensity spectra. Using 3D simulations, \citet{georgobiani03}  computed mode line profiles in both the velocity and the intensity power spectrum alternatively at mean unity optical depth and instantaneous unity optical depth. Their results indeed show that the modulation of the `observed'\ intensity fluctuations due to radiative transfer could be significant enough to reverse the sense of mode asymmetry. One of the hypothesis enjoying the most support for asymmetry reversal, however, is based on the effect of turbulent perturbations partially correlated with the mode, which thus impact its line profile \citep{nigamCORREL98, roxburghV97, rastB98, kumarB99}. Indeed, a part of these perturbations is coherent with the mode and, thus, leads to interference. This interference term may be constructive or destructive, depending on the phase difference between the mode and the coherent turbulent perturbations. For frequencies at which the interference is constructive, the power spectral density is slightly elevated, whereas it drops slightly for frequencies at which it is destructive. Typically, in the vicinity of a resonant mode, the dependence of the phase difference between mode and turbulent perturbation is such that the interference term is constructive for frequencies located in one wing of the mode and destructive in the other. Therefore, as a result of this interference behaviour, one of the wings falls off more slowly and the other more rapidly, leading to mode asymmetry. It has been suggested that the degree of correlation between the turbulent perturbations and the oscillation it excites is higher in intensity than in velocity, so that it changes the sign of mode asymmetry only in the intensity spectrum. While it is widely accepted that correlated turbulent perturbations must be taken into account to explain asymmetries in the intensity spectrum, the question of whether it has a significant impact on the velocity spectrum remains an open issue \citep[see e.g.  ][]{jefferies03}.

The possibility that correlated turbulent fluctuations have an affect on mode asymmetry has led many authors to include them in their models for the power spectrum. Even though correlated noise was introduced to explain the particular puzzle of asymmetry reversal between velocity and intensity measurements, several models include correlated noise in the velocity spectrum as well as in the intensity spectrum. This is the case for the model developed by \citet{severino01} and later used, for instance, by \citet{barban04}, which includes three types of noise (coherent-correlated, coherent-uncorrelated and incoherent) in both the velocity spectrum, the intensity spectrum, and the velocity-intensity cross-spectrum. They considered, however, that the `pure oscillation' (without the noise) has a Lorentzian shape, thus discarding the contribution of source localisation. This model was later refined by \citet{watcherK05} to take this contribution into account.

These prior studies have one thing in common, however, and that is that they all treat the various sources of asymmetry (mainly source localisation and correlated noise) in a simplified, parametrised fashion. Indeed, the excitation is consistently modelled as a point-like source, with radial position and multipolar development left as free parameters. This prescription remains somewhat unsatisfactory in the sense that it does not take into account the finer properties of the source of excitation, such as its spatial extent or its dependence on frequency, for instance. As such, these prior models lack a realistic description of the source of excitation. Likewise, for models including the effect of noise on the power spectrum, the various relative amplitudes and phase differences between modal oscillation and correlated noise in both spectra are also left as free parameters. The approach followed by these studies is to find best-fit values for all their free parameters by fitting their model to observations in order to localise the source.

In contrast, in the present paper, we follow a different approach: we model both the source of excitation and the correlated background by constraining their properties using an analytical model of stochastic excitation, coupled with a 3D simulation of the solar atmosphere. The novelty of our approach lies in the fact that we do not fit a parametrised model to the observations but, instead, we predict the dependence of mode asymmetry on frequency, which we then compare to observations in order to validate our model. Our model of mode asymmetry is, therefore, both more realistic (in its description of the source of excitation) and more complete (in its lack of freely adjustable parameters). It can then be used to deepen our understanding of the underlying physical mechanisms behind mode asymmetry by exploring how varying a given physical constraint impacts the results yielded by our model. Finally, our model allows for a much higher predictability of mode asymmetry, which is essential when it comes to applying these results to other solar-like oscillators. We note that this paper is devoted to the modelling of the velocity power spectrum only and, as a result, we do not address the problem of asymmetry reversal, which is a separate challenge altogether.

These efforts to model the line profiles of solar-like oscillations are also necessary in order to correctly infer mode properties from observations. Indeed, it was discovered early on that using a Lorentzian shape to fit skew symmetric line profiles led to a significant bias in the eigenfrequency determination, which may be higher than the frequency resolution achieved by helioseismic measurements \citep{duvall93, abramsK96, chaplin99, thiery00, toutain98}. Such eigenfrequency determination bias has also been revealed for solar-like oscillations in stars other than the Sun by \citet{benomar18}. Inversion methods used to infer the internal structure of solar-like oscillators, whether they be analytical or numerical, require a very accurate determination of the mode eigenfrequencies. For spectra extracted from very long time series, the resolution is high enough that this bias in eigenfrequencies impacts the results obtained by inversion methods \citep[see e.g.   ][who show that the difference between the sound speed squared inferred from symmetric and asymmetric fits can reach $0.3 \%$ in the core]{toutain98}. When fitting these observations, mode asymmetry must, therefore, be taken into account. Since it has proven very difficult to determine accurate mode eigenfrequency without prior knowledge on its line shape, obtaining an a priori model of $p$-mode line profiles is of primary importance.

In this paper, we present a predictive model of solar radial $p$-mode line profile in the velocity spectrum. In particular, we use a realistic model for stochastic excitation, following a method similar to that of \citet{samadiG01}. Furthermore, we include the effect of correlated turbulent perturbations in the model in a non-parametrised way, unlike what was done in previous works \citep[see e.g. ][]{severino01}. The paper is structured as follows: we present the analytical model of the Sun's velocity power spectral density in Section \ref{sec:methods} and its numerical implementation in Section \ref{sec:numerical}. We then present the results yielded by our model concerning mode asymmetry in Section \ref{sec:results}. In Section \ref{sec:discussion}, we briefly describe the development of a toy model to describe the impact of source localisation on mode asymmetry and use it to interpret our results; we also investigate the matter of the influence of correlated turbulent perturbations. We then confront our results with the related observations in Section \ref{sec:observations} and discuss the issue of eigenfrequency determination bias entailed by the skewness of the mode line profiles.

\section{Modelling the $p$-mode line profiles}\label{sec:methods}

To extract the asymmetries of solar radial $p$-modes, we first need to model their line profile in the velocity power spectrum. In this section, we present the analytical developments that led us to this model. First, we define the disc-integrated velocity power spectrum in terms of the radial fluid displacement. We then present the inhomogeneous, radial wave equation associated to the acoustic modes and detail how convolving its Green's function with its inhomogeneous part gives us access to the velocity power spectral density.

\subsection{Definition of the velocity power spectral density}

Before embarking on a discussion of the actual modelling of the line profiles, the spectrum from which they are extracted needs to be defined. In this paper, we restrict ourselves to the study of radial acoustic modes in the Sun. Furthermore, as part of the definition of the spectrum, we include the effect of limb-darkening and of disk integration that affect the Sun-as-a-star measurements. We note, however, that other instrumental effects - in particular mode leakage - are not accounted for.

To derive an expression for the observed power spectral density, we separate the total surface velocity into an oscillatory part $\bm{\vv_\text{osc}}$ and a turbulent part $\bm{u}$, where it is understood that the modes are described by the oscillatory part. The observations made for the Sun-as-a-star are obtained by integrating the velocities over the entire solar disk. Neither the mode velocity (for radial modes), nor the turbulent perturbations depend on the point of the disk at which it is estimated; however, the projection on the line of sight $\bm{n}$ does. This integration over the solar disk is affected by limb-darkening $h(\mu)$ (where $\mu$ refers to the cosine of the angle between the local radial direction and the line of sight). Furthermore, since it is the turbulent perturbations that excite the mode, a certain fraction of the former must be correlated with the latter, so that the contribution of turbulent perturbations to the velocity spectrum must be considered.

With these considerations, the observed velocity power spectral density can be expressed as

\begin{equation}
P(\omega) = \dfrac{1}{\displaystyle\int \diff\Omega~h(\mu)}\left\langle \left|\displaystyle\int \diff\Omega~h(\mu)\left(\widehat{\bm{\vv_\text{osc}}}(\omega) + \widehat{\bm{u}}(\omega)\right).\bm{n} \right|^2 \right\rangle ,
\label{eq1}
\end{equation}
where the integration is performed over the solar disk, $\Omega$ refers to the solid angle, $\bm{n}$ is the unit vector along the line of sight, $\bm{\vv}_\text{osc}$ is the mode velocity, $\bm{u}$ represents the fluctuations of the turbulent velocity around its mean value, $\omega$ is the angular frequency, the notation $\left(~\widehat{.}~\right)$ refers to temporal Fourier transform, and $\langle . \rangle$ refers to ensemble average. Since we are only considering radial modes, $\bm{\vv_\text{osc}}$ is exclusively radial. Thus, Eq. \eqref{eq1} becomes

\begin{equation}
P(\omega) = \left\langle \left|\widehat{\vv_\text{osc}}(\omega)\displaystyle\int \diff\Omega~\mu \widetilde{h}(\mu) + \displaystyle\int \diff\Omega~\widetilde{h}(\mu)\widehat{u_n}(\omega) \right|^2 \right\rangle ,
\end{equation}
where $u_n$ is the component of the turbulent velocity along the line of sight. We introduced the reduced limb-darkening $\widetilde{h}(\mu)$ so that its integral over the solar disk is normalised to unity. 

We expand the square in the above expression and we consider that the term containing $\langle|\widehat{u_n}|^2\rangle$ is negligible compared to the terms that contain $\langle|\widehat{\vv_\text{osc}}|^2\rangle$ and $\mathrm{Re}\left(\langle \widehat{u_n} \widehat{\vv_\text{osc}}^\star\rangle\right)$ , respectively. Indeed, the power spectral density is several orders of magnitude higher for the mode velocity than for the turbulent velocity \citep[typically, the former is of order $10^5$ m$^2$.s$^{-2}$.Hz$^{-1}$, while the latter is of order $10$ m$^2$.s$^{-2}$.Hz$^{-1}$, e.g.][Fig. 2]{turck04}, so that
\begin{equation}
\langle|\widehat{u_n}|^2\rangle \ll \mathrm{Re}\left(\langle \widehat{u_n} \widehat{\vv_\text{osc}}^\star\rangle\right) \ll \langle|\widehat{\vv_\text{osc}}|^2\rangle~,
\end{equation}
where the notation $\mathrm{Re}$ refers to the real part of a complex quantity, and $^\star$ refers to its complex conjugate. Finally, we obtain

\begin{multline}
P(\omega) = \left(\displaystyle\int \diff\Omega~\mu~\widetilde{h}(\mu)\right)^2 \left\langle \left|\widehat{\vv_\text{osc}}(\omega)\right|^2 \right\rangle \\
+ 2\displaystyle\int \diff\Omega~\mu~\widetilde{h}(\mu) \text{Re}\left(\displaystyle\int \diff\Omega~\widetilde{h}(\mu) \left\langle \widehat{\vv_\text{osc}}(\omega)\widehat{u_n}^{\star}(\omega) \right\rangle\right) .
\label{eq:dev}
\end{multline}

The first term corresponds to the spectral power density of the mode velocity $\vv_\text{osc}$. In itself, the line profile generated by this term is already asymmetric; indeed, it has been known for a long time that source localisation can generate line profile asymmetry \citep[see  e.g.][]{abramsK96, roxburghV97, chaplinA99}. The second term corresponds to what the literature commonly refers to as correlated turbulent perturbations and which is also expected to significantly impact mode asymmetry in photometric measurements \citep[see e.g.][]{nigamCORREL98, roxburghV97, kumarB99}, although its importance in velocity measurements is not as clear.

\subsection{The inhomogeneous wave equation}

Going further, we write the radial wave equation associated to $\vv_\text{osc}$ with the same formalism as \citet{unno89}. We detail its derivation in Appendix \ref{app:wave_eq}. Although we included both the source terms due to Reynolds stress fluctuations and non-adiabatic pressure fluctuations in the computation detailed in Appendix \ref{app:wave_eq}, we only consider the former  in the following. Indeed, it is the dominant source of excitation for acoustic modes in the Sun \citep[e.g.][]{belkacem08}. When it is temporally Fourier transformed, the inhomogeneous wave equation for radial modes reads:

\begin{equation}
\dfrac{\diff^2\Psi_\omega}{\diff r^2} + \left(\dfrac{\omega^2}{c^2} - V(r)\right)\Psi_\omega = S(r)~,
\label{eq2}
\end{equation}
where $c$ is the sound speed, the wave variable $\Psi_\omega(r)$ is related to the radial fluid displacement $\xi_r(r)$ through

\begin{equation}
\Psi_\omega(r) = rc(r)\sqrt{\rho_0(r)}\xi_r(r)~,
\label{eq:psi_to_v}
\end{equation}
and the acoustic potential and source term are given by
\begin{equation}
\begin{array}{l}
V(r) = \dfrac{N^2-4\pi G\rho_0}{c^2} + \dfrac{2}{x^2}\left(\dfrac{\diff x}{\diff r}\right)^2 - \dfrac{1}{x}\dfrac{\diff^2x}{\diff r^2}~,\\
x(r) = \dfrac{r\sqrt{I}}{c}~, \\
S(r) = \dfrac{r}{c\sqrt{\rho_0(r)}} \dfrac{\diff p_t'}{\diff r}~, \\
I(r) = \exp\left(\displaystyle\int_0^r \dfrac{N^2}{g_0}-\dfrac{g_0}{c^2}~\diff r'\right)~,
\end{array}
\label{eq:potential_mainbody}
\end{equation}
where $r$ is the radial coordinate, $\rho_0$ is the density, $N$ is the Brunt-Väisälä frequency, $g_0$ is the gravitational acceleration, $G$ is the gravitational constant, and $p_t'$ refers to the fluctuations of the Reynolds stress around its mean value. Indeed, only the fluctuating part of the Reynolds stress contributes to the source term $S(r)$ and its mean value only modifies the equilibrium structure. The subscript $0$ refers to the equilibrium structure and all the above quantities are dependent on the radius at which they are estimated, even when not explicitly specified. We note that we only model radial modes in this paper, so that the wave equation (Eq. \ref{eq2}) is of the second order despite the fact that we did not use the Cowling approximation.

Mode damping is not included in Eq. \eqref{eq2}. Indeed, we did not take into account the feedback of modal oscillations on the equilibrium state through modulations in the fluid density, pressure, opacity, etc. Such feedback allows mechanical work and thermal transfer to occur from the mode to the medium in which it develops; depending on the phase-lag between these different modulations energy can be exchanged with the surrounding medium. However, the modelling of damping rates of solar-like oscillations is extremely difficult \citep{samadiBS15}. Thus, we directly introduce damping in the wave equation in the form of a mode lifetime, or, equivalently, by a (frequency-dependent) linewidth $\Gamma_\omega$, so that the wave equation takes the following form

\begin{equation}
\dfrac{\diff^2\Psi_\omega}{\diff r^2} + \left(\dfrac{\omega^2 + j\omega\Gamma_\omega}{c^2} - V(r)\right)\Psi_\omega = S(r) ,
\label{eq:wave_equation}
\end{equation}
where $j$ denotes the imaginary unit and the linewidths $\Gamma_\omega$ are inferred from observations. We used the line-widths presented in \citet{baudin05} (see their Table 2), which were inferred from GOLF data. Note, however, that our definition of $\Gamma_\omega$ corresponds to their $\Gamma$ multiplied by $2\pi$, or equivalently to twice their damping rate $\eta$. We completed these data with low-frequency line-widths obtained by \citet{davies14} through BiSON, which go as low as $\sim 900~\mu$Hz (see their Table 1). We reproduce the dependence of the linewidth we used on frequency in Table \ref{table:gammas}. We note that damping can potentially be a source of mode asymmetry. However, the impact of damping on mode asymmetry is very weak compared to the other sources of asymmetry \citep{abramsK96}, so that the direct introduction of observed line-widths in our model is unlikely to have an impact on our results.

\begin{table}
\centering
\begin{tabular}{cc|cc}
$\nu$ ($\mu$Hz) & $\Gamma_\omega$ ($\mu$Hz) & $\nu$ ($\mu$Hz) & $\Gamma_\omega$ ($\mu$Hz) \\
\hline\hline & \\
\vspace{-20pt} & \\
972.615 & 0.0055 & 2828.15 & 0.94 \\
1117.993 & 0.0091 & 2963.29 & 0.80 \\
1263.198 & 0.022 & 3098.16 & 1.08 \\
1407.472 & 0.033 & 3233.13 & 1.12 \\
1548.336 & 0.082 & 3368.56 & 1.84 \\
1686.594 & 0.20 & 3504.07 & 2.83 \\
1749.33 & 0.26 & 3640.39 & 3.85 \\
1885.10 & 0.28 & 3776.61 & 5.90 \\
2020.83 & 0.47 & 3913.49 & 8.09 \\
2156.79 & 0.54 & 4049.46 & 10.73 \\
2292.03 & 0.74 & 4186.98 & 12.69 \\
2425.57 & 0.88 & 4324.79 & 16.39 \\
2559.24 & 0.94 & 4462.08 & 17.35 \\
2693.39 & 0.92 & 4599.96 & 26.42
\end{tabular}
\caption{Observational linewidth $\Gamma_\omega$ used in Eq. \eqref{eq:wave_equation} as a function of frequency $\nu$. The data are extracted from \citet{baudin05} for frequencies higher than $1750~\mu$Hz, and from \citet{davies14} below. When a frequency laid between these points, linear interpolation was used.}
\label{table:gammas}
\end{table}

\subsection{Expression of the velocity power spectral density}

By definition, the Green's function $G_\omega(r_\text{o}, r_\text{s})$ is the value taken by the function $\Psi_\omega$ at the radius $r = r_\text{o}$ (the variable $r_\text{o}$ refers to the height in the atmosphere at which the spectrum is observed and the variable $r_\text{s}$ refers to the position of the point-like source term), where $\Psi_\omega$ is the solution to the inhomogeneous wave equation,

\begin{equation}
\dfrac{\diff^2\Psi_\omega}{\diff r^2} + \left(\dfrac{\omega^2 + j\omega\Gamma_\omega}{c^2} - V(r)\right)\Psi_\omega = \delta(r - r_\text{s})~,
\label{eq:avant}
\end{equation}
and $\delta$ refers to the Dirac function. Once the Green's function is known, it can be used to express explicitly $\vv_\text{osc}$ in Eq. \eqref{eq:dev}. Indeed, on the one hand, the general solution to the inhomogeneous wave equation with a source term $S(r_\text{s})$ is

\begin{equation}
\Psi_\omega(r_\text{o}) = \displaystyle\int \diff r_\text{s}~G_\omega(r_\text{o}, r_\text{s}) S(r_\text{s})~,
\label{eq:convolv}
\end{equation}
where the source term is given by Eq. \eqref{eq:potential_mainbody}. The pulsational velocity $\vv_\text{osc}$ is related to the variable $\Psi_\omega$ through

\begin{equation}
\vv_\text{osc}(r_\text{o}) = \dfrac{j\omega}{r_\text{o}c(r_\text{o})\sqrt{\rho_0(r_\text{o})}}\Psi_\omega(r_\text{o})~.
\label{eq:puls_vel}
\end{equation}

Using the source term given by Eq. \eqref{eq:potential_mainbody} in Eq. \eqref{eq:convolv}, and Eq. \eqref{eq:puls_vel} and after finally performing an integration by part, we write the velocity Fourier transform at angular frequency $\omega$ as

\begin{multline}
\widehat{\vv_\text{osc}}(\omega, r_\text{o}) = -\dfrac{j\omega}{r_\text{o} c(r_\text{o})\sqrt{\rho_0(r_\text{o})}} \\
\times \displaystyle\int \diff^{3}\bm{r_\text{s}}~\bm{\nabla}\left( G_\omega(\bm{r_\text{s}}, r_\text{o})\dfrac{||\bm{r_\text{s}}||}{c(\bm{r_\text{s}})\sqrt{\rho_0(\bm{r_\text{s}})}}\right) . \left(\rho_0 \widehat{u_r \bm{u}})(\bm{r_\text{s}})\right) .
\label{eq:v}
\end{multline}

In the following, the observation height $r_\text{o}$ will be fixed, so that we drop it for ease of notation. However, since the observation height depends on the transition line used for the observations and on whether the observations rely on spectrometric or photometric measurements, it significantly varies from instrument to instrument (see Sect. \ref{sec:observations} for more details).

Using Eq. \eqref{eq:v} in Eq.\eqref{eq:dev} then gives an expression for the velocity power spectral density in terms of Green's function $G_\omega(r_\text{s}):$

\begin{equation}
P(\omega) = \left(\displaystyle\int \diff\Omega~\mu~\widetilde{h}(\mu)\right)^2 \left[\left\langle \left|\widehat{\vv_\text{osc}}(\omega)\right|^2 \right\rangle + C(\omega)\right] ,
\label{eq:power}
\end{equation}
where $\left\langle \left|\widehat{\vv_\text{osc}}(\omega)\right|^2 \right\rangle$ and $C(\omega)$ are given, respectively, by Eqs. \eqref{eq:dominant}, and \eqref{eq:crossed}. We note that the effects of limb-darkening and disk integration are now contained in a single factor and, thus, these will only have an effect on mode amplitude. Since the asymmetry of a mode does not depend on its amplitude, it is not impacted by such a factor.

The calculations leading from Eq. \eqref{eq:dev} to Eq. \eqref{eq:power} are detailed in Appendix \ref{app:spectrum}. In the following, we only provide the main steps and assumptions. We split the calculations two ways, focussing separately on the first term inside the brackets of Eq. \eqref{eq:power}, which we hereby refer to as the leading term, and on its second term, which we hereby refer to as the cross term. 

\subsubsection{Closure models\label{subsec:moments}}

The calculations leading from Eq. \eqref{eq:dev} to Eq. \eqref{eq:power} involve the evaluation of fourth-order and third-order two-point correlation moments of the turbulent velocity. Therefore, an appropriate closure model is needed to express these high-order moments as a function of second-order moments. We devote the following subsection to presenting and developing these closure models.

\paragraph{Fourth-order moments}

To describe the fourth-order correlation moments of the turbulent velocity, we make use of the Quasi-Normal Approximation (QNA hereafter). This closure model consists in considering that all turbulent quantities are normally distributed, in which case fourth-order moments can be analytically expressed as a combination of second-order moments \citep{book_lesieur}:\ 

\begin{equation}
\langle abcd \rangle = \langle ab \rangle \langle cd \rangle + \langle ac \rangle \langle bd \rangle + \langle ad \rangle \langle bc \rangle ,
\label{eq:decomposition}
\end{equation}
where $a$, $b$, $c$ , and $d$ refer to any turbulent scalar quantity. Applying the QNA to isotropic, homogeneous turbulence inhibits energy transfers among modes of different wave numbers, thus leading to violations of the energy conservation principle \citep{kraichnan57}. This is due to the fact the QNA entails vanishing third-order correlation moments. When it comes to estimating the fourth-order moments, however, the picture is different. \citet{belkacem06B} have studied the validity of the QNA for two-points, fourth-order correlation moments of the vertical turbulent velocity, in the form of $\langle u_{r,1}^2 u_{r,2}^2 \rangle$ (where the indices $1$ and $2$ refer to two different points in space), using 3D simulations of the solar atmosphere. They found that the dependence of this correlation moment on the distance $\Delta X$ between the two points is correctly estimated by the QNA but that its absolute value (which can be taken as the corresponding one-point moment) is not. Consequently, the amplitude of the modes are largely underestimated when the QNA is used. However, the asymmetry of the modes does not depend on their amplitude, so that mode asymmetry will be unaffected by a discrepancy in the absolute value of the two-points, fourth-order moments. As such, the decomposition given by Eq. \eqref{eq:decomposition} can be considered valid when it comes to studying mode asymmetry.

\paragraph{Third-order moments}

While the QNA provides an adequate closure relation for fourth-order moments, as mentioned above, it assumes vanishing third-order moments. Therefore, in order to estimate these third-order moments, we make use of another closure model, the \textit{Plume closure model} (PCM hereafter), which was developed by \citet{belkacem06}. The idea behind this closure model is to separate the flows directed upwards from those directed downwards (the latter being referred to as plumes) and to apply the QNA to both separately. The anisotropy between the two types of flow - in particular, turbulence is more prominent in the downwards plumes \citep[e.g.][]{goode98} - yields non-vanishing third-order correlation moments:

\begin{multline}
\langle u_r(\bm{R},t)^2 u_r(\bm{R}+\bm{r},t+\tau)\rangle = \left[a(1-a)^3 - a^3(1-a)\right] \delta u^3 \\
-a(1-a)\left[2\langle \widetilde{u_d}(\bm{R},t) \widetilde{u_d}(\bm{R}+\bm{r},t+\tau) \rangle + \langle \widetilde{u_d}(\bm{R},t)^2 \rangle\right]\delta u ,
\label{eq:PCM}
\end{multline}
where $u_r$ is the vertical component of the turbulent velocity, $a$ is the relative horizontal section of the upflows, $\delta u$ is the difference between the mean velocity of the two types of flows (considering their respective signs, it actually is the sum of their absolute values), and $\widetilde{u_d}$ is the fluctuation of the vertical velocity around its mean value in the downflows.

We note that, strictly speaking, the third-order moment given by Eq. \eqref{eq:PCM} and yielded by the PCM are centred. However, we consider that the mean value of the overall vertical velocity of the flow is sufficiently low (compared to its standard deviation for instance) to be neglected. Therefore, these moments may interchangeably refer here either to centred or non-centred moments.

We also note that this closure relation is written here in terms of $\widetilde{u_d}$ (i.e. the turbulent fluctuations in the downflows only). It would be more practical to rewrite it in terms of $u_r$ (i.e. the total turbulent fluctuations). The two are related through
\begin{equation}
\langle \widetilde{u_d}(\bm{R}, t)\widetilde{u_d}(\bm{R}+\bm{r}, t+\tau) \rangle = \dfrac{1}{1-a} \langle u_r(\bm{R}, t)u_r(\bm{R}+\bm{r}, t+\tau) \rangle - a\delta u^2~.
\end{equation}

\subsubsection{The leading term\label{subsec:leading}}

In the following, we detail the derivation of the first term of Eq. \eqref{eq:power}. This term corresponds to the pulsational velocity itself, without correlated turbulent perturbations. As such, any asymmetry featured by this term alone represents the effect of source localisation. The first step consists in separating the scales relevant to the turbulent velocity $\bm{u}$ from the scales relevant to both the medium stratification and the oscillating mode (respectively, the pressure scale height and the mode wavelength). The scale separation approximation is not realistic in the subsurface layers (in particular, the mode wavelength is comparable to the typical correlation length associated with turbulence); however, for want of a better alternative, we are led to use this approximation in the following.

Since the integral defining $\widehat{\vv_\text{osc}}(\omega)$ in Eq. \eqref{eq:v} contains the turbulent velocity fluctuations squared, expanding the square of its modulus will raise these fluctuations to the fourth. The contribution of turbulence to the expression of $\vv_\text{osc}$ thus takes the form of two-points, fourth-order correlation moments of the turbulent velocity. We use the closure relation presented and detailed in Subsection \ref{subsec:moments} to express them as a function of second-order moments.

We then use analytical expressions for the second-order moments of the turbulent velocity. We describe the second-order moment of the $i$-th and $j$-th component of the turbulent velocity in terms of its spatial and temporal Fourier transform $\phi_{ij}(\bm{k},\omega)$. For isotropic turbulence, it reads \citep{batchelor_book}:
\begin{equation}
\phi_{ij}(\bm{k},\omega) = \dfrac{E(k,\omega)}{4\pi k^2} \left(\delta_{ij} - \dfrac{k_i k_j}{k^2}\right)~,
\end{equation}
where $E(k,\omega)$ is the specific turbulent kinetic energy spectrum, $k$ is the norm of the wavevector $\bm{k}$, $k_i$ and $k_j$ are its $i$-th and $j$-th component, and $\delta_{ij}$ is the Kronecker symbol. The integration over the solid angle of wave vectors $\bm{k}$ is straightforward, and only an integral over the norm of $\bm{k}$ remains. However, solar turbulence close to the photosphere is known to be highly anisotropic. To take this anisotropy into account, we follow the formalism developed by \citet{gough77}. In this formalism, the integral over the solid angle of $\bm{k}$ is simply readjusted by adding an anisotropy factor given by Eq. \eqref{eq:anisotropy} \citep[see Appendix B in][]{samadiG01}.

Following \citet{stein67}, we then decompose $E(k,\omega)$ into a spatial part $E(k)$, which describes how the turbulent kinetic energy is distributed among modes of different wave numbers, and a temporal part $\chi_k(\omega)$, which describes the statistical distribution of the life-time of eddies of wavenumber $k$

\begin{equation}
E(k,\omega) = E(k)\chi_k(\omega)~.
\end{equation}

In order to model the spatial and temporal part of the spectrum of turbulent kinetic energy, we followed two different approaches, described in the following.

\paragraph{The `theoretical spectrum' model\label{para:theory_spec}}

We use theoretical prescriptions to model both the spatial spectrum $E(k)$ and the temporal spectrum $\chi_k(\omega)$ of turbulent velocity. Based on the assumption that turbulent flows are self-similar, Kolmogorov's theory of turbulence leads to a spatial spectrum $E(k) \propto k^{-5/3}$ in the inertial range, between $k = k_0$ (where $k_0$ is the scale at which the kinetic energy is injected in the turbulent cascade, and is henceforth referred to as the injection scale) and the dissipation scale (at which the turbulent kinetic energy is converted into heat). Given the very high Reynolds number characterising solar turbulence ($\mathrm{Re} \sim 10^{14}$), we cast the dissipation scale to infinity. Then, following \citet{musielak94}, we extend the turbulent spectrum below the injection scale by considering that $E(k)$ takes a constant value for $k < k_0$. This extended spectrum, referred to as the broadened Kolmogorov spectrum (BKS hereafter) was introduced to account for the broadness of the maximum of $E(k)$. The BKS can be written as

\begin{equation}
E(k) = \left\{
\begin{array}{ll}
0.652\dfrac{u_0^2}{k_0} & \text{ if } 0.2~k_0 < k < k_0 \\
0.652\dfrac{u_0^2}{k_0}\left(\dfrac{k}{k_0}\right)^{-5/3} & \text{ if } k_0 < k ,
\end{array}
\right.
\end{equation}
where $u_0^2 \equiv \langle \bm{u}^2(\bm{r}) \rangle / 3$ and the $0.652$ factor is introduced so that the total specific kinetic energy of the turbulent spectrum matches $u_0^2 / 2$. Therefore, the spatial spectrum is parametrised solely by the injection scale $k_0$. However, the injection scale varies significantly between the sub-surface layers and the atmosphere \citep{samadi03}, so that we keep it free in our model and allow for it to depend on the radial coordinate $r$.

Following \citet{samadi03}, we consider a Lorentzian shape for the temporal spectrum $\chi_k(\omega)$, which is supported both by numerical simulations \citep{samadi03} and by theoretical arguments. Indeed, a noise described by a stationary, Gaussian Markov process in time is expected to relax exponentially, meaning that the resulting eddy-time correlation is expected to be a decreasing exponential, and its Fourier transform a Lorentzian function \citep{belkacem11levrai}. The width $\omega_k$ associated to eddy-time correlation is linked to the life-time of the eddies of wavenumber $k$. Dimensional arguments would suggest that $\omega_k \propto k u_k$, where $u_k$ is the typical velocity associated to the eddies of wavenumber $k$. However, there remains a substantial indetermination on the actual value of $\omega_k$, so that, following \citet{balmforth92}, we consider:

\begin{equation}
\omega_k = 2k u_k / \lambda~,
\label{eq:lambda}
\end{equation}
where $\lambda$ is a dimensionless, constant parameter. Overall, the only input parameters of this model are $k_0(r)$ and $\lambda$.

\paragraph{The `numerical spectrum' model}

In the second model, we extract the spatial spectrum $E(k)$ from a 3D hydrodynamic simulation of the solar atmosphere, provided by the $\text{CO}^5\text{BOLD}$ code (see Sect. \ref{sec:numerical} for details). This simulation gives us access to the velocity field as a function of all three spatial coordinates and time. In order to extract the turbulent spectrum $E(k)$, we average the velocity field temporally, then isolate each horizontal slice in the simulated cube and perform a 2D Fourier transform of each slice separately, thereof which we only retain the radial part. This gives us a spectrum $E(k)$ for each vertical point in the simulation. Finally, we renormalise each spectrum so that

\begin{equation}
\displaystyle\int_0^{+\infty} \diff k~E(k) = \dfrac{u_0^2}{2}~,
\end{equation}
where $u_0$ is also extracted from the 3D atmospheric simulation, by averaging the fluid velocity squared temporally and horizontally, and using the definition $u_0^2 = \langle \bm{u}^2(\bm{r}) \rangle / 3$.

The temporal spectrum $\chi_k(\omega)$ is also treated in a slightly different manner than in the `theoretical model' above. Indeed, the arguments invoked above to justify the Lorentzian shape of the spectrum, while valid for most of the relevant time scales associated to turbulent eddies, are no longer valid for shorter time scales, that is, for higher angular frequencies. \citet{belkacem11} argued that if the time correlation associated to small eddies indeed originates from their advection by larger, energy-bearing eddies - a hypothesis referred to as the  sweeping assumption -  one recovers a Gaussian spectrum instead of a Lorentzian one. The transition between a Lorentzian spectrum, valid for low angular frequencies, and a Gaussian spectrum, valid for high angular frequencies, occurs at the cut-off angular frequency $\omega_E$, which is given by the curvature of the eddy-time correlation function at $\tau=0$ \citep{belkacem11}:
\begin{equation}
\omega_E = k u_0~.
\end{equation}

Since a Gaussian spectrum would fall off much more rapidly than a Lorentzian spectrum, we simply consider that $\chi_k$ vanishes entirely for $\omega > \omega_E,$

\begin{equation}
\chi_k(\omega) =
\left\{\begin{array}{ll}
\dfrac{1}{2\arctan(\omega_E / \omega_k) \omega_k}\dfrac{1}{1+(\omega/\omega_k)^2} & \text{if}~\omega < \omega_E \\
0 & \text{if}~\omega_E < \omega~.
\end{array}
\right.
\end{equation}

We modified the prefactor so that $\chi_k$ meets the normalisation condition. The typical life time of eddies of wavenumber $k$, parametrised by $\omega_k$, is still given by Eq. \eqref{eq:lambda}. We note that the convolution of the function $\chi_k(\omega)$ with itself must be computed to evaluate the leading term (see Eq. \ref{eq:must_convol}). While the convolution of a Lorentzian function with itself straightforwardly yields a Lorentzian function with a width twice as large, the convolution of the modified spectrum above with itself is slightly different, but can be obtained analytically as

\begin{multline}
(\chi_k \ast \chi_k)(\omega) = \dfrac{1}{2\pi\omega_k} \dfrac{1}{1+(\omega/2\omega_k)^2} \\
\times \dfrac{\pi\left(\arctan\left(\dfrac{\omega_E}{\omega_k}\right) - \arctan\left(\dfrac{\omega-\omega_E}{\omega_k}\right)\right)}{4\arctan^2\left(\dfrac{\omega_E}{\omega_k}\right).}
\end{multline}

Physically, taking the cut-off frequency into account significantly decreases the predicted amplitudes for high frequency modes. As far as mode asymmetry is concerned, we found that it did not have a significant impact in the `theoretical spectrum' model. In contrast, it substantially changes mode asymmetry at high frequency in the `numerical spectrum' model, which is why we only introduce it in the latter. The reason is the following: the spatial spectrum of turbulent energy falls off much more rapidly with $k$ in the `theoretical spectrum' than in the `numerical spectrum'. Therefore, small spatial scales play a more important role in the latter. Since the cut-off frequency only impacts these small scales, it is natural that taking it into account should only have a significant impact on the `numerical spectrum' model.

\subsubsection{The cross term}

The derivation of the cross term follows essentially the same steps as for the leading term. The main difference is that the turbulent velocity correlation moments that appear are now two-point, third-order moments. We use the closure relation presented in Sect. \ref{subsec:moments} to express them, as we did with the fourth-order moments, as a function of two-point, second-order moments of the turbulent velocity. We then use the same analytical description of second-order moments as the one we used for the leading term. The rest of the calculations is very similar to those described in Sect. \ref{subsec:leading} and leads to the second term in Eq. \eqref{eq:power}.

These two models - the `theoretical spectrum' and `numerical spectrum' models - are complementary in the sense that the first one allows us to explore the impact of the properties of turbulence on mode asymmetries and gives physical insight into this problem, whereas the second one relies on fewer input parameters and, therefore, has more predictive capability (we recall here that the former requires the parameters $\lambda$ and $k_0(r)$ to be set, while the latter only requires $\lambda$). Consequently, in the following, we present and develop the results yielded by both.

\section{Numerical implementation}\label{sec:numerical}

In this section, we detail how we numerically implemented the model presented in Sect. \ref{sec:methods}. We describe how we obtained the solar equilibrium state in which the acoustic modes develop and how we integrated the inhomogeneous wave equation given by Eq. \eqref{eq:wave_equation}. Having obtained the solar radial $p$-mode line profiles, we then detail how we extracted their asymmetries and perform several tests to validate our model and its numerical implementation.

\subsection{The solar equilibrium state \label{subsec:eq_state}}

The acoustic potential given by Eq. \eqref{eq:potential_mainbody} depends only on the equilibrium structure of the Sun. We extracted the potential from a 1D solar model provided by the evolutionary code CESTAM \citep{morel97, marques13}. The 1D model includes treatment of the convective flux (using standard mixing-length theory with no overshoot) and of the radiative flux (using the Eddington approximation). On the other hand, turbulent pressure, rotation, and diffusion processes are ignored.

However, 1D stellar models do not fully take into account the more complex physical phenomena taking place in the uppermost layers of a star; in particular, the rapid transition between the convective zone and the superficial radiative region \citep{kupkaM17}. This leads to significant biases in the equilibrium structure. Since the excitation of solar oscillations precisely takes place in these layers, it is essential that we model them more accurately. To do so, we use a 3D hydrodynamic simulation of the solar atmosphere computed using the $\text{CO}^5\text{BOLD}$ code \citep{freytag12}. The modelled region includes the super-adiabatic peak just below the photosphere and goes up to the lower atmosphere of the star.

\begin{figure}
\centering
\includegraphics[width=\linewidth, trim=20 0 0 0]{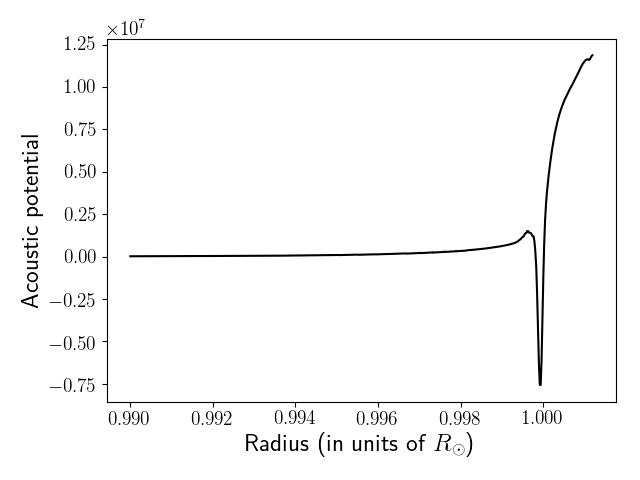}
\caption{Acoustic potential $V(r)$ used in Eq. \eqref{eq:wave_equation}, calculated using Eq. \eqref{eq:potential_mainbody} and the equilibrium state of the Sun given by the solar patched model described in the text. The radius is normalised by the photospheric radius $R_\odot$, and only the outermost region is shown. We note that the acoustic potential is normalised by $R_\odot^{-2}$ here (where $R_\odot$ is the radius of the solar photosphere) so that it is dimensionless.}
\label{fig:potential}
\end{figure}

It is now possible to construct a `patched' model\ of the solar interior. We use the solar patched model computed by \citet{manchon}. The process of constructing patched models has been extensively discussed \citep[e.g.][]{trampedach97, samadi08} and the particular case of the patched model used in this paper is described in much detail in \citet{manchon}. The basic idea is to transform the 3D atmosphere into a 1D atmosphere through temporal and horizontal averaging and then to replace the surface layers of the 1D stellar interior with this 1D atmosphere. We note that the input parameters of the CESTAM model used to describe the solar interior (age, total stellar mass, mixing-length parameter $\alpha_\text{MLT}$ , and helium abundance) are chosen so that the top layers match the bottom layers of the CO$^5$BOLD atmosphere. Here the model was computed with the mixing-length parameter $\alpha_\text{MLT} = 1.65$, an initial helium abundance of $Y_\text{init} = 0.249,$ and an initial metallicity of $Z_\text{init} = 0.0135$. Figure \ref{fig:potential} shows the acoustic potential profile $V(r)$ given by this solar patched model, computed using Eq. \eqref{eq:potential_mainbody}.

Finally, we use the same simulation of the solar atmosphere to extract the various parameters appearing in the analytical description of the source term; in particular, the standard deviation $u_0$ associated to turbulent velocities, the anisotropy factor $G$ given by Eq. \eqref{eq:anisotropy}, as well as the parameters used in the \textit{Plume closure model} (see Eq. \ref{eq:PCM}). Specifically, the ensemble average appearing in the definition of $u_0$ was computed by performing a temporal and horizontal average of the norm of the velocity squared in the 3D simulation of the solar atmosphere. We also used this same simulation to extract the spatial spectrum of turbulent kinetic energy in the `numerical spectrum' model (see Sect. \ref{sec:methods}).

\subsection{Integration of the inhomogeneous wave equation}

To compute one value of $P(\omega)$ for one value of the angular frequency $\omega$ (i.e. one point in the velocity power spectrum), we convolve the Green's function $G_\omega(r_\text{s})$ associated to Eq. \eqref{eq:wave_equation} with the stochastic source term $S(r_\text{s})$ (see Sect. \ref{sec:methods}). It is then possible to reconstruct the velocity power spectral density, and in particular the line profile of the resonant modes, point by point (typically, we only need 10 points regularly spaced between $\omega_0 - \Gamma_{\omega_0}$ and $\omega_0 + \Gamma_{\omega_0}$, where $\omega_0$ is the angular eigenfrequency of the mode and $\Gamma_{\omega_0}$ its linewidth). In the following, we describe how the wave equation was integrated and how we extracted its Green's function.

For a given angular frequency $\omega$, we carry out the integration using a 4th-order Runge-Kutta scheme \citep{press_book} with the acoustic potential $V(r)$ given by the solar equilibrium state (see Sect. \ref{subsec:eq_state}). Given that radial modes develop in the entire solar volume, we perform this integration over the entire solar radius, between $r = 0$ and $r = r_\text{max}$. We note that $r_\text{max}$ refers not to the photospheric radius, but to the maximum radial extent of the solar model described in Sect. \ref{subsec:eq_state}, so that $r_\text{max} > r_\text{photosphere}$.

We imposed Dirichlet boundary conditions on the wave equation. The condition at the centre is straightforward: by definition, $\Psi_\omega(r=0) = 0$. At $r_\text{max}$, we impose a vanishing Lagrangian pressure perturbation (which physically means that the atmosphere of the Sun is force-free). The continuity equation and the equation of state allow us to rewrite this latter condition in terms of $\Psi_\omega:\ $

\begin{equation}
\dfrac{\diff\Psi_\omega}{\diff r} + \dfrac{\diff}{\diff r}\left[\ln\left(\dfrac{r}{c\sqrt{\rho_0}}\right)\right] \Psi_\omega = 0~.
\end{equation}

The use of Dirichlet boundary conditions leads us to implement a shooting method: we perform the integration with $\Psi_\omega(r=0) = 0$, and tune the initial slope (i.e. the value of $\diff\Psi_\omega / \diff r$ at $r=0$) until the other boundary condition is met. Note that this method is not similar to the shooting method usually implemented to solve the eigenvalue problem associated to the determination of acoustic mode eigenfrequencies: here, the pulsation $\omega$ is fixed, and it is the initial slope that is tuned to meet the boundary condition at the surface. The difference between these two methods is that in the inhomogeneous problem, the initial slope (or, alternatively, the amplitude of the mode) is fixed by the amplitude of the source of excitation. The shape of the eigenfunction, however, remains the same as in the homogeneous problem.

This method enables us to extract the Green's function associated to the wave equation (Eq. \ref{eq:wave_equation}). To obtain one value of the Green's function $G_\omega(r_\text{s})$, for one value of the pulsation $\omega$ and one value of the source position $r_\text{s}$, we carry out the integration of the inhomogeneous equation as described above, adding a point-like source term to the numerical scheme. The source is normalised in such a way that the right-hand side equals $1/h$ when the source falls within the integration radial step, and $0$ otherwise ($h$ is the radial step of the integration).

This integration gives us the radial oscillation profile $\Psi_\omega(r)$, and we simply extract its value at a fixed radius $r_\text{o}$, which corresponds to the height in the atmosphere at which the spectrum is measured. We note that the presence of damping in the wave equation implies that it is complex-valued. As such, the Green's function is complex-valued as well.

Finally, to calculate the integrals over source positions which appear in Eqs. \eqref{eq:dominant} and \eqref{eq:crossed}, we compute the Green's function using the above method for a grid of source positions $r_\text{s}$, while $\omega$ is kept constant. This grid corresponds to the radial grid provided by the 3D atmospheric model described in Sect. \ref{subsec:eq_state}.

We also use the aforementioned 3D model to extract the physical quantities appearing in both the leading term of Eq. \eqref{eq:power} (the anisotropy factor $G$, the turbulent velocity fluctuations $u_0$), and the cross term (the parameters $a$ and $\delta w$ in the PCM; see Sect. \ref{sec:methods} for a definition of these parameters).

Using Eq. \eqref{eq:power} provided by the model developed in Sect. \ref{sec:methods} and the radial grid of Green's functions computed using the above method finally allows us to extract the value of $P(\omega)$ for each value of $\omega$.

\subsection{Fitting of the mode asymmetries}

\begin{figure}
\centering
\includegraphics[width=\linewidth, trim=20 0 0 0]{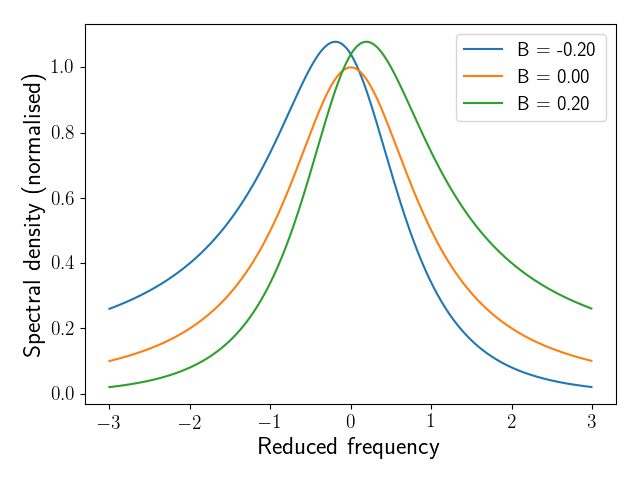}
\caption{Dependence of the line profile given by Eq. \eqref{eq:formula} on the asymmetry parameter $B$. The line profiles are normalised with $H_0 = 1$.}
\label{fig:illustrB}
\end{figure}

We fit the line profile of the modes following \citet{nigamFORMUL98} with the formula

\begin{equation}
P(\omega) = H_0\dfrac{(1+Bx)^2 + B^2}{1+x^2} ,
\label{eq:formula}
\end{equation}
where $x = 2(\omega - \omega_0)/\Gamma_{\omega_0}$ is the reduced pulsation frequency. The fit contains three free parameters ($H_0$, $\nu_0$ and $B$), the last of which is defined as the asymmetry parameter. We illustrate the dependence of the line profile on $B$ in Fig. \ref{fig:illustrB}. In particular, $B < 0$ means that the peak contains more power in the low-frequency side (that corresponds to negative asymmetry), $B > 0$ means that the high-frequency side contains more power (that corresponds to positive asymmetry), while with $B=0$ we recover a classic, Lorentzian profile. Figure \ref{fig:illustrB} also shows that the mode does not peak exactly at the eigenfrequency, but rather at a slightly higher (for $B > 0$) or lower value (for $B < 0$). This can have important repercussions for the determination of the mode eigenfrequencies from observations, as we discuss in Section \ref{sec:bias}.

Note that several other definitions of the asymmetry parameters can be found in the literature. \citet{korzennik05} prefers to adjust the mode line profiles with

\begin{equation}
P(\omega) \propto \dfrac{1 + \alpha(x - \alpha / 2)}{x^2 + 1}~,
\end{equation}
and defines the asymmetry parameter as $\alpha_{n,l}$. Meanwhile, \citet{vorontsovJ13} use the following formula:

\begin{equation}
P = H\left[\left(\dfrac{A\cos(\phi - S)}{1 - R^2}\right)^2 + B^2\right]~,
\end{equation}
where the frequency variable is $\phi$, and the asymmetry parameter is defined as $S$. While these three formulas have been derived in different ways, they are perfectly equivalent close to the eigenfrequency (for $x \ll 1$, or equivalently for $\phi \equiv 0 \pmod \pi$), with $S \sim B \sim \alpha / 2$.

Finally, \citet{gizon06} provides the asymmetry parameter defined as \citep[see also][]{benomar18}:
\begin{equation}
\chi = 2B\omega_0 / \Gamma_\omega~.
\label{eq:chi}
\end{equation}

The author quantified the mode asymmetry by means of the relative positions of the local maxima (peaks) and minima (troughs) in the power spectral density: minima located half-way between the neighbouring maxima lead to symmetric line profiles, while a deviation from this behaviour leads to asymmetric line profiles. The parameter $\chi$ derived from these considerations is independent from both the amplitude and the line-widths of the modes.

The formulas presented above only lead to different line profiles far from the central frequency, whereas they are equivalent in our range of interest. We opted for the definition given by \citet{nigamFORMUL98} (Eq. \ref{eq:formula}) because it is the most commonly used.

To ensure the significance of fitting an asymmetric profile to the mode obtained through our model, we compared the results produced by the fitting formula Eq. \eqref{eq:formula} and by a symmetric, Lorentzian profile (that is, imposing $B = 0$ in Eq. \ref{eq:formula}). The asymmetric fits led to excellent agreement with the modelled line profiles; however, the symmetric fits led to substantial discrepancies, with one wing consistently falling off more rapidly than the numerical line profile and the other too slowly. Finally, it should be noted that the excellent fit given by Eq. \eqref{eq:formula} to the numerical line profile is independent from the number of points used for the adjustment; we have indeed performed a similar fit with thrice the number of points, without any loss of accuracy and the resulting asymmetry parameter $B$ was the same to an excellent approximation.

\subsection{Validation of the method}

Using the method presented above, we extracted solar radial modes of radial orders $n=6$ to $n=30$. Indeed, the formula used for the fit and given by Eq. \eqref{eq:formula} does not converge properly for higher-order modes (because the increasing linewidths lead to mode overlapping), while we did not have access to observed linewidths for lower-order modes. In addition to their line profile asymmetries, we also extract other fundamental properties, namely their eigenfrequencies, amplitudes, and eigenfunctions. In the following, to support the validity of our model, we compare the mode properties we obtained with similar properties obtained through other methods.

First, we compare the eigenfrequencies obtained through our model to the eigenfrequencies of the 1D adiabatic oscillations calculated using the ADIPLS code \citep{jcd11}. For this validation, we did not make use of the patched model described in Sec. \eqref{subsec:eq_state} but, rather, the corresponding unpatched model. The reason is that the patching procedure produces a small discontinuity of the physical quantities at the patching point, which can affect the eigenfrequencies calculated by ADIPLS. We recover the correct eigenfrequencies, with errors not exceeding $\sim 0.1 \%$. Since mode asymmetry is only expected to vary on the scale of $\sim$ mHz, modelled asymmetries will not be significantly affected by such small discrepancies of the eigenfrequencies.

\begin{figure}
\centering
\includegraphics[width=\linewidth, trim=20 0 0 0]{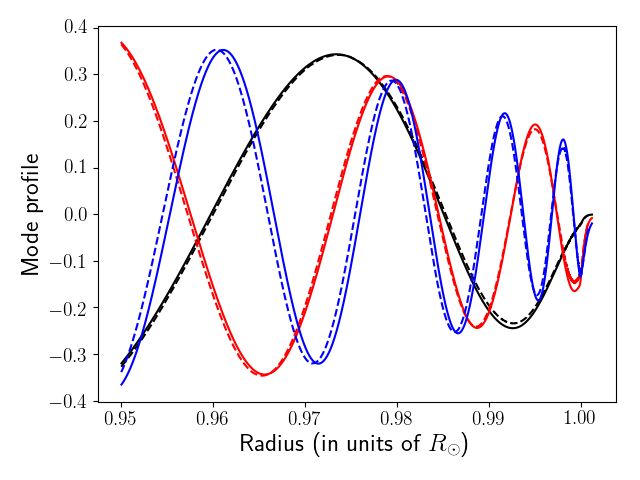}
\caption{Eigenfunction $\Psi(r)$ of several radial acoustic modes (\textit{black}: $n=10$ ($\nu_0 = 1.580$ mHz); \textit{red}: $n=20$ ($\nu_0 = 2.912$ mHz); \textit{blue}: $n=30$ ($\nu_0 = 4.267$ mHz)) as computed by our model (\textit{solid lines}), and calculated using the ADIPLS code (\textit{dashed lines}). The radial axis is zoomed to show only the outermost region. The eigenfunctions have been normalised so that their maximum over the entire solar interior equals $1$.}
\label{fig:profiles}
\end{figure}

Our numerical method also allows us to extract the radial profile $\Psi(r)$ of the eigenmodes. We compare them in Fig. \ref{fig:profiles} to the eigenfunctions calculated using the same 1D adiabatic oscillations obtained through the ADIPLS code and presented above. The figure shows that the modes obtained through our model have eigenfunctions that are very similar to those obtained through this dedicated code, which further supports the validity of the model we have used.

Finally, we compare the mode amplitudes obtained through our model to the observed ones. To that end, we estimated the velocity power spectrum at an observation height of $340$ km, which corresponds to the observation height of the GOLF instrument as estimated by \cite{baudin05}, following \cite{brulsR92}. By definition, the velocity amplitude squared is the total area under the mode peak, so that it depends both on its maximum $H$ and on its width $\Gamma,$

\begin{equation}
\vv_\text{osc} = \sqrt{\pi H\Gamma}~.
\label{eq:height_to_amp}
\end{equation}
We note that when it is used to treat observational data, this formula also contains a geometric factor to account for instrumental effects, including mode visibility. This factor is, however, irrelevant in our case.

\begin{figure}
\centering
\includegraphics[scale=0.6]{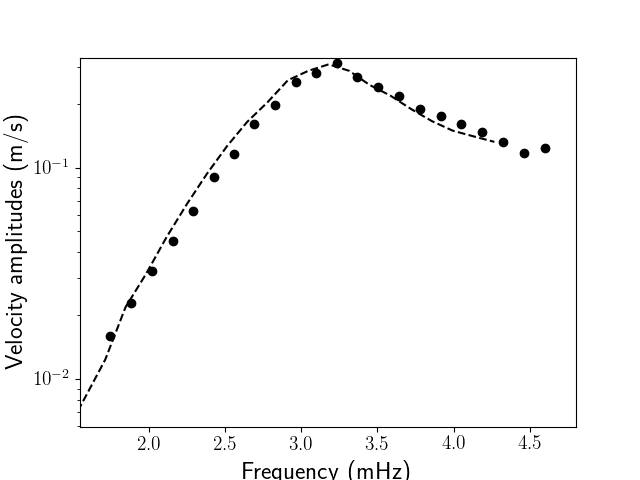}
\caption{Velocity amplitudes of radial acoustic modes as computed by our model, using Eq. \eqref{eq:height_to_amp} (\textit{black dashed line}), and as observed by GOLF (\textit{black points}). The data points are taken from \citet{baudin05}. The free parameters in the model have been tuned to obtain the best possible agreement with observational data.}
\label{fig:heights}
\end{figure}

We show in Fig. \ref{fig:heights} the comparison between the mode amplitudes $\vv_\text{osc}$ obtained through our `numerical spectrum' model and the mode amplitudes inferred from observations performed by the GOLF instrument \citep{baudin05}. The free parameter $\lambda$ of our model (\textit{cf.} Sect. \ref{sec:methods}) has been adjusted so as to obtain the best possible agreement. As a consequence, our model does not hold any predictive power when it comes to mode amplitudes. However, the fact that we manage to retrieve a very good agreement with observational data by using reasonable values of the input parameters is still a solid sign that our model is valid. In particular, we correctly recover the frequency at maximum amplitude $\nu_\text{max}$, as well as the slopes on both the low-frequency and the high frequency limit. To conclude on the matter, we emphasise that the asymmetry parameter $B$ is independent from the mode amplitude, so that potential discrepancies concerning the latter should not affect the former.

\section{Results for the asymmetry profile $B(\nu)$}\label{sec:results}

Using the model presented in Sect. \ref{sec:methods}, numerically implemented using the method presented in Sect. \ref{sec:numerical}, we extract the solar $p$-modes line profile asymmetries $B(\nu)$ throughout a large part of the spectrum, between $n=6$ ($\nu \sim 1$ mHz) and $n=30$ ($\nu \sim 4.3$ mHz). In this section, we present the results yielded by our model, focusing on the dependence of the asymmetry parameter $B$ on frequency (which we hereafter shorten to the expression `asymmetry profile') and on the impact of our different input parameters on the asymmetry profile.

As we detailed in Sect. \ref{sec:methods}, we followed two different approaches to model the turbulent kinetic energy spectrum. The first one, which we refer to as the `theoretical spectrum' model, uses the prescription given by Kolmogorov's theory of turbulence and which we have described in detail in Sect. \ref{sec:methods}. The second approach, which we refer to as the `numerical spectrum' model, uses the turbulent spectrum extracted from the 3D hydrodynamic simulation of the solar atmosphere described in Sect. \ref{sec:numerical}. In this section, we present separately the asymmetry profile $B(\nu)$ yielded by both models.

\subsection{The `theoretical spectrum' model\label{subsec:theoretical_spec}}

This model relies on a prescription for the properties of turbulence. It contains the following input parameters: the temporal spectrum of turbulent kinetic energy, parametrised by the dimensionless quantity $\lambda$, which is defined by Eq. \eqref{eq:lambda}, and its spatial spectrum, parametrised by $k_0(r)$, which is defined as the (radius-dependent) injection wavenumber of turbulent kinetic energy. We let the latter depend on $r$ in order to account for the fact that the typical size of turbulent eddies drastically depends on where they are located with respect to the photosphere. It is known that the size of the energy-bearing eddies increases with height, so that the injection scale $k_0$ decreases with $r$ \citep{samadi03}. We simplify the situation by considering that the injection rate only takes two values: $k_0(r) = k_{0,\text{int}}$ below the photosphere, and $k_0(r) = k_{0,\text{atm}}$ above the photosphere. This picture crudely corresponds to what is observed in 3D atmospheric simulations \citep{samadi03}. In the following, we denote the ratio between the two as $R_k \equiv k_{0,\text{int}} / k_{0,\text{atm}}$. This leaves only three input parameters in our model: $\lambda$, $k_{0,\text{int}}$ , and $k_{0,\text{atm}}$; or equivalently $\lambda$, $k_{0,\text{int}}$ , and $R_k$.

\begin{figure}
\centering
\includegraphics[width=\linewidth, trim=20 0 0 0]{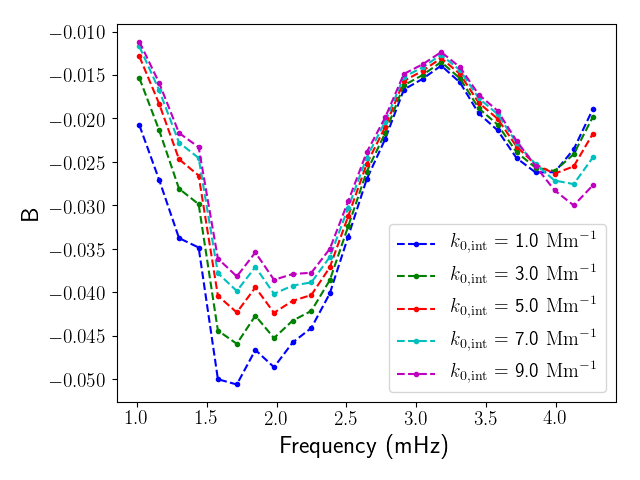}
\caption{Asymmetry parameter $B$ as a function of frequency obtained by the `theoretical spectrum' model, for different values of $k_{0,\text{int}}$, with $\lambda = 1$ and $R_k = 2$ fixed. The sub-photospheric injection scale $k_{0,\text{int}}$ is expressed in $\text{Mm}^{-1}$.}
\label{fig:asym_vs_k0abs}
\end{figure}

\begin{figure}
\centering
\includegraphics[width=\linewidth, trim=20 0 0 0]{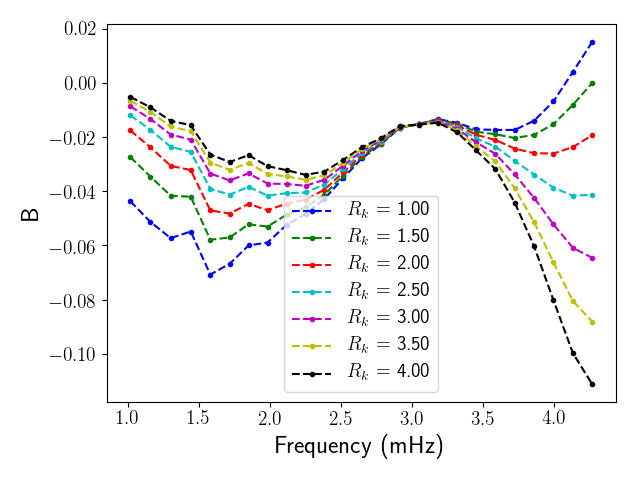}
\caption{Same as Fig. \ref{fig:asym_vs_k0abs}, but only $R_k$ varies, $\lambda = 1$ and $k_{0,\text{int}} = 2\text{ Mm}^{-1}$.}
\label{fig:asym_vs_rats}
\end{figure}

\begin{figure}
\centering
\includegraphics[width=\linewidth, trim=20 0 0 0]{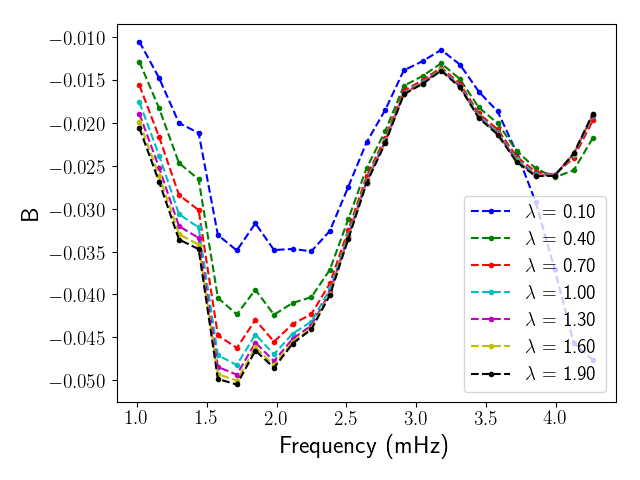}
\caption{Same as Fig. \ref{fig:asym_vs_k0abs}, but only $\lambda$ varies, $R_k = 2$ and $k_{0,\text{int}} = 2\text{ Mm}^{-1}$.}
\label{fig:asym_vs_lambdas}
\end{figure}

In Fig. \ref{fig:asym_vs_k0abs}, we keep $\lambda$ and $R_k$ constant, and we show the asymmetry profile $B(\nu)$ for several values of $k_{0,\text{int}}$. Despite the fact that we vary $k_{0,\text{int}}$ across almost one order of magnitude, the asymmetry profile $B(\nu)$ does not depend significantly on the absolute value of $k_0$, except close to $\nu \sim 1.7$ mHz. By comparison, its dependence on both $R_k$ and $\lambda$ is more substantial, especially at high frequencies (\textit{cf.} Figs. \ref{fig:asym_vs_rats} and \ref{fig:asym_vs_lambdas}). Since $k_{0,\text{int}}$ does not seem to play an important role, we keep it fixed in the following, and focus on the impact of the other two input parameters, $\lambda$ and $R_k$.

Fig. \ref{fig:asym_vs_k0abs} illustrates the main qualitative features of the asymmetry profile $B(\nu)$. In fact, together with Figs. \ref{fig:asym_vs_rats} and \ref{fig:asym_vs_lambdas}, it shows that the qualitative behaviour of the asymmetry profile is largely model-independent. Thus the asymmetries of the solar radial $p$-mode line profiles are negative across a large part of the spectrum, in agreement with solar observations \citep[see for instance][]{duvall93}. Furthermore, the asymmetry profile $B(\nu)$ exhibits three distinct local extrema: the absolute value of $B$ increases below $\sim 1.7$ mHz, decreases between $\sim 1.7$ and $\sim 3$ mHz, increases again between $\sim 3$ mHz and $\sim 4$ mHz, and finally decreases again above $\sim 4$ mHz. Note, however, that this last extremum is, unlike the other ones, somewhat impacted by the values given to the different input parameters of the model.

The first two local extrema ($\sim 1.7$ and $\sim 3$ mHz) correspond respectively to the beginning and end of the damping rate plateau. Indeed, the asymmetry parameter $B$ depends on the linewidth of the modes, so it is natural that a sudden change in the behaviour of the latter should reflect on the former. The third extrema is not so easily explained and it will be discussed in Sect. \ref{sec:discussion}.

Fig. \ref{fig:asym_vs_rats} shows how the asymmetry profile $B(\nu)$ depends on $R_k$. An increase of this parameter attenuates low-frequency mode asymmetries (below $\nu_\text{max} \sim 3$ mHz), while on the contrary, the high-frequency modes (above $\nu_\text{max}$) become more asymmetric. The effect is significantly more substantial for the latter than for the former. Asymmetries close to $\nu_\text{max}$, however, are not affected by the parameter $R_k$ whatsoever.

Figure \ref{fig:asym_vs_lambdas} shows that the impact of $\lambda$ on the asymmetry profile $B(\nu)$ is quite similar, albeit inverted, in the sense that $|B|$ increases with $\lambda$ for low-frequency modes and decreases for high-frequency modes. Similarly, $B$ is barely impacted by a change of $\lambda$ close to $\nu_\text{max}$. Furthermore, the asymmetry profile $B(\nu)$ undergoes saturation, in the sense that it ceases to depend on $\lambda$ when it is increased above a certain value. In the following, we denote this threshold as $\lambda_\text{sat}$. Figure \ref{fig:asym_vs_lambdas} shows that $\lambda_\text{sat} \sim 1$. This dichotomy between $\lambda \lesssim 1$ and $\lambda \gtrsim 1$ originates in the Lorentzian nature of the temporal turbulent spectrum: depending on the value of $\lambda$, the angular frequencies relevant to solar $p$-modes are either in the low frequency part or in the high frequency part of the spectrum. We do not go into too much detail here as we discuss this matter further in Sect. \ref{sec:discussion}.

\subsection{The `numerical spectrum' model}

In the `numerical spectrum' model, which describes the properties of turbulence more realistically, there is only one input parameter left, $\lambda$. In this sense, it has a greater predictive power than the previous model. The qualitative behaviour of the asymmetry profile $B(\nu)$ and, in particular, the positions of the different local extrema featured by $B(\nu)$, are, in this model, rather independent from $\lambda$ and in agreement with what we observed in the scope of the previous model.

\begin{figure}
\includegraphics[width=\linewidth, trim=20 0 0 0]{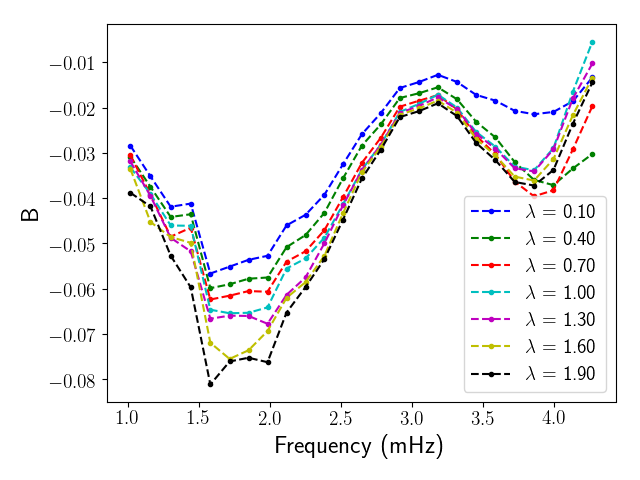}
\caption{Asymmetry parameter $B$ as a function of frequency obtained by the `numerical spectrum' model, for several values of $\lambda$.}
\label{fig:asym_vs_lambda_newmodel}
\end{figure}

However, the input parameter $\lambda$ does have an impact on the quantitative behaviour of the asymmetry profile $B(\nu)$. We show in Fig.\ref{fig:asym_vs_lambda_newmodel} the asymmetry profile $B(\nu)$ obtained with the `numerical spectrum' model (see Sect. \ref{sec:methods}) for several values of $\lambda$. As for the `theoretical spectrum' model, $B$ is always negative and features several local extrema at $\nu \sim 1.7$, $3$ and $4$ mHz.

As for the dependence of $B(\nu)$ on $\lambda$, two distinct regimes can be separated. Below $\nu_\text{max} \sim 3$ mHz, we recover the same dependence of the asymmetry parameter $B$ with $\lambda$ as we obtained in the scope of the `theoretical spectrum' model, with absolute values of $B$ increasing with $\lambda$. The picture at frequencies higher than $\nu_\text{max}$ is, however, somewhat different. The asymmetry profile $B(\nu)$ features a local minimum at $\nu \sim 4$ mHz; the curve inflexion grows sharper as $\lambda$ increases up to $\lambda \sim 1$, after which this part of the asymmetry profile does not significantly depend on $\lambda$. In this sense, the asymmetry profile $B(\nu)$ seems to undergo the same saturation behaviour as in the `theoretical spectrum' model (see Sect. \ref{subsec:theoretical_spec}), for the same value $\lambda_\text{sat} \sim 1$. The fact that we recover approximately the same threshold gives us confidence that this particular feature of the asymmetry profile $B(\nu)$ is not a mere artefact of one model or the other but, rather, it is a genuine effect based on a physical origin. Again, we postpone the discussion of the physical origin of this behaviour to Sect. \ref{sec:discussion}.

\section{Impact of the properties of turbulence on mode asymmetry}\label{sec:discussion}

Line profile asymmetry of solar-like oscillations have two main causes: localisation of the source of excitation \citep[see for instance][]{duvall93} and correlation with the turbulent perturbations \citep[see for instance][]{nigamCORREL98}. In the following, we investigate both contributions in light of the results yielded by the `theoretical spectrum' model and presented in Sect. \ref{subsec:theoretical_spec}. With its various input parameters, the `theoretical spectrum' model allows us to understand the physical origin of mode asymmetry. In this section, then we only consider this model, although the conclusions are valid for the `numerical spectrum' model as well. We first discuss how source localisation and correlated turbulent perturbations can skew the mode line profiles. In particular, we support the discussion concerning source localisation with a simplified toy-model of mode excitation, which we describe in Appendix \ref{app:toy_model}. We then use this discussion to interpret the results yielded by our model. Additionally, we show that the contribution of the correlated turbulent perturbations to the mode asymmetries is negligible in the velocity power spectrum.

\subsection{Origin of mode asymmetry}

\subsubsection{Effect of source localisation on mode asymmetry: generic arguments}

The fact that the source of excitation of a mode is spatially localised can affect the skewness of the mode line profile in Fourier space. There are several ways of describing the impact of source localisation on mode asymmetry.

One way is to make use of the analogy between the development of acoustic modes in the stellar cavity and the phenomenon of optical interference in a Fabry-Pérot cavity. This analogy was used to account for the acoustic mode asymmetry in the Sun by \citet{gabriel92, duvall93}, among others. The idea is that acoustic, stationary modes in the Sun can be described by means of two progressing waves, propagating in opposite directions. Each of these waves follows the same cycle: they propagate one way, get refracted on the lower turning point of the acoustic cavity, then propagate backwards, get reflected on the upper turning point, and so forth. As a result of these multiple reflections and refractions on both turning points, the acoustic waves pass multiple times through the same regions and, therefore, interfere with each other (and with themselves). This interference pattern leads to the development of resonant modes in the cavity. What we observe then is the evanescent tail of these modes in the atmosphere, which lies outside the resonant cavity.

Let us now consider that the source of the waves is located at a certain point within the cavity. The waves propagating outwards and inwards will have travelled over different distances before interfering with one another and this difference of travel times will depend on the location of the source. The shape of the mode line profile is directly related to the dependence of the phase difference between the outwards and inwards interfering waves on frequency. Since this phase difference is not exactly symmetric about the mode eigenfrequency, neither is its line profile; and given that it depends on the source location, mode asymmetry is indeed a marker of source localisation.

Another physical interpretation of how source localisation can bring about mode asymmetry has been proposed by \citet{rastB98}, and later refined by \citet{rosenthal98}. They remarked that mode asymmetry could be mathematically described by the relative position of local maxima (or peaks) and local minima (or troughs) in the power spectrum. Peaks located exactly halfway between their neighbouring troughs feature symmetric, Lorentzian line profiles. However, if one of the neighbouring trough is closer than the other, the peak in question appears skewed and, depending on which trough is closest, its asymmetry parameter is either positive or negative.

The position of the peaks are simply related to the eigenmodes of the solar acoustic cavity. As for the position of the troughs, in the special case of a point-like source of excitation, with a given multipolar decomposition, the authors showed that it is related to the eigenmodes of the atmosphere truncated at the source position, with a vanishing external boundary condition depending on the multipolar nature of the source. In that interpretation, the position of the troughs thus depends on both the position and the multipolar decomposition of the source.

Yet another way to describe the impact of source localisation on mode asymmetry is to consider the eigenfunction of the mode. In order to illustrate this, we present in Appendix \ref{app:toy_model} a very simplified toy-model of mode excitation, where the source is considered point-like and the acoustic cavity is simplified to a square well potential. From this toy-model we draw the following conclusion: for a given frequency, the amplitude of the wave is proportional to the eigenfunction associated with the wave at the source of excitation. In particular, excitation at a mode's antinode is much more efficient than at a mode's node.

\begin{figure}
\includegraphics[width=\linewidth]{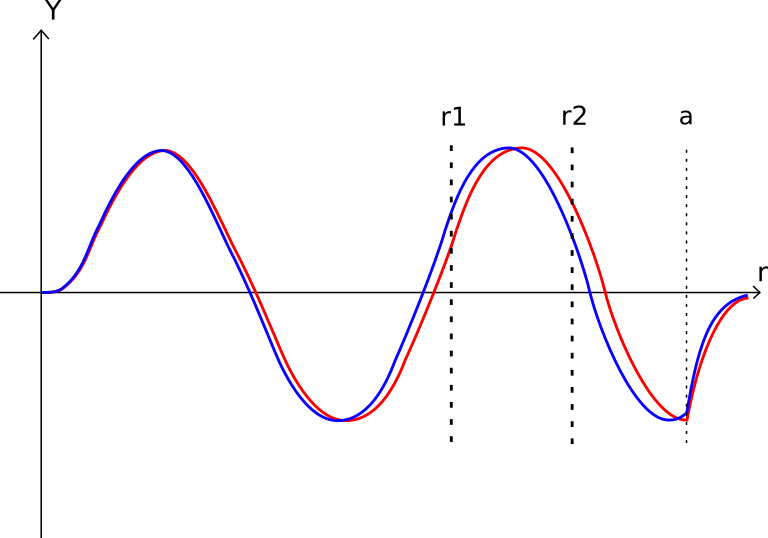}
\caption{Illustration of the importance of source position with respect to nodes and antinodes of the eigenfunction associated to a mode to explain its asymmetry. The blue and red lines show the radial profile of the oscillation for two angular frequencies very close to one another ($\omega_\text{red} < \omega_\text{blue}$). The bold vertical dashed lines show two source positions generating opposite mode asymmetries: positive for $r_1$, negative for $r_2$. The third vertical dashed line marks the edge of the acoustic cavity $r=a$.}
\label{fig:schema}
\end{figure}

With this conclusion in mind, let us consider the situation illustrated by Fig. \ref{fig:schema}. The blue and red curves represent the radial profile of the acoustic wave for two different angular frequencies. It can be seen that an increase of $\omega$ causes the radial profile of the oscillation to `shrink' radially. Therefore, the amplitude of the oscillation as seen by the source will either increase or decrease with $\omega$, depending on its position. More specifically, a source at $r = r_1$ (see illustration in Fig. \ref{fig:schema}) will see the amplitude of the oscillation increase with $\omega$, and a source at $r = r_2$ will see it decrease. In light of the conclusion presented in the previous paragraph, it can be deduced that if the source is located at $r = r_1$, the right wing of the mode line profile will be slightly elevated compared to the left wing, thus leading to positive asymmetry. Likewise, the asymmetry generated by a source at $r = r_2$ will be negative.

From the illustration in Fig. \ref{fig:schema}, it is straightforward to see that the dichotomy between the $r = r_1$ case and the $r = r_2$ case is based on the relative position of the source and the nodes and antinodes of the mode, or, in other words, on the sign of the derivative of the absolute value of the eigenfunction. To be more specific, one has to separate the case of a source inside and outside the acoustic cavity. If the source is inside the cavity, the $r = r_1$ case (i.e. case where source localisation entails positive asymmetry) corresponds to any source position located above a node and below an antinode of the oscillation profile, whereas the $r = r_2$ case (i.e. the case where source localisation entails negative asymmetry) corresponds to any source position located above an antinode and below a node. Here we recall  that a node is a point at which the wave amplitude is zero and an antinode is a point at which it is maximal. If the source is outside the cavity, however, it is always as in the $r = r_2$ case and, thus, it always generates negative asymmetry: indeed, the outside of the cavity corresponds to an evanescent zone for the acoustic waves so that the absolute value of the eigenfunction always decreases in this region.

It should be noted that we only consider this toy-model in the present subsection. In the following sections, we return to the discussion of our model, simply using the conclusions drawn above to interpret the results which it yields.

\subsubsection{Correlated turbulent fluctuations}

Acoustic modes in the Sun are excited by fluctuations of turbulent nature - more specifically by turbulent fluctuations of the Reynolds stress or non-adiabatic pressure perturbations. It is therefore natural that a part of the turbulent fluctuations should be not only coherent, but statistically correlated with the oscillating mode.

The resulting interference between the mode and the turbulent fluctuations leads, in turn, to mode asymmetry. In order to illustrate this, let us consider a mode whose line profile is intrinsically Lorentzian and turbulent fluctuations whose power spectral density is constant over the width of the mode under consideration. We then have
\begin{equation}
P(x) = \left|\dfrac{A_m}{x+j} + A_n e^{j\phi_n}\right|^2 ,
\end{equation}
where $P$ is the total power spectral density, $x = 2(\omega-\omega_0)/\Gamma_\omega$ is the reduced frequency ($\omega_0$ is the angular eigenfrequency of the mode, and $\Gamma_\omega$ its linewidth), $A_m$ and $A_n$ are the (real) amplitudes associated to the mode and the noise respectively, $\phi_n$ is the phase difference between the mode and the noise, and $j$ is the imaginary unit. Expanding the module squared, we obtain
\begin{equation}
P(x) = \dfrac{A_m^2}{1+x^2} + \dfrac{2A_m A_n}{\sqrt{1+x^2}}\sin\left(\arctan x + \phi_n\right) + A_n^2 .
\label{eq:correlated_noise}
\end{equation}

The first term of the right-hand side of Eq. \eqref{eq:correlated_noise} corresponds to a Lorentzian profile and is symmetric about $x = 0$. The third term simply acts as an offset and does not introduce any mode asymmetry. The second term, however, is clearly not symmetric at $x = 0$, unless $\phi_n = \pm \pi / 2$. For instance, if $\phi_n = 0$, this term is even antisymmetric. In other words, the interference between the mode and the noise is destructive in the left wing of the mode and constructive in its right wing. As such, the power spectral density $P(x)$ is higher than the Lorentzian profile in the right wing and lower in the left wing, thus entailing positive mode asymmetry. The sign and magnitude of the mode asymmetry depends on the amplitude $A_n$ , that is, on the degree of correlation between the mode and the turbulent fluctuations, as well as on the phase difference $\phi_n$, both of which are included in the model we developed in Sect. \ref{sec:methods}.

\subsection{Contribution of source localisation to $B(\nu)$}

In the previous subsection, we summarised the impact of source localisation on mode asymmetry by stating the following: a source within the resonant cavity of a mode entails negative asymmetry if it is located above an antinode and below a node of the associated eigenfunction and positive asymmetry otherwise; a source outside the resonant cavity always entails negative mode asymmetry. With this in mind, we set out to interpret the results obtained in Sect. \ref{sec:results} in the scope of the `theoretical spectrum' model.

Once applied to the case of solar $p$-mode excitation, this rule can be rephrased in the following way. There is a dichotomy between the effect of the turbulent eddies located below the upper turning point of the mode and those located above. The former skew the mode line profile one way or the other depending on their height relative to the nodes and antinodes of the mode eigenfunction. The latter always skew the mode line profile so that it feature negative asymmetry. Until now, we have only discussed the case of a point-like source of excitation. However, the driving region of the solar $p$-modes, while localised around the super-adiabatic peak just below the photosphere, has a certain spatial extent. As such, driving turbulent eddies can be found both below and above the upper turning point of the modes and the observed mode asymmetry is due to the combination of both.

\begin{figure}
\includegraphics[width=\linewidth, trim=20 0 0 0]{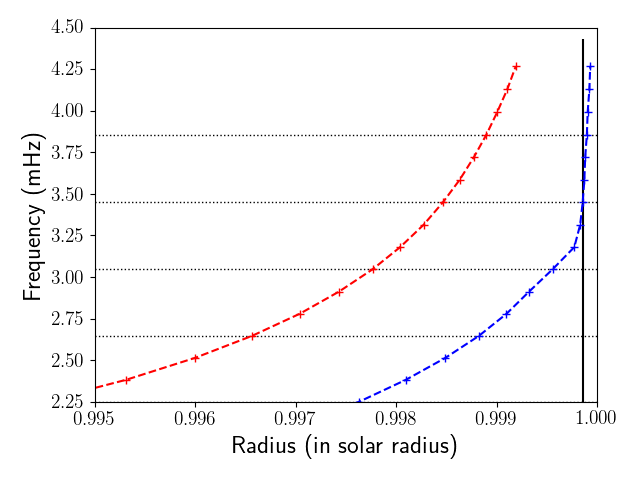}
\caption{Radial location of the nodes (\textit{red symbols connected by a dashed line}) and anti-nodes (\textit{blue symbols connected by a dashed line}) that are closest to the super-adiabatic peak, for each radial mode between $n=16$ and $n=30$. The vertical black line represents the maximum of the super-adiabatic peak, where the excitation is most efficient. Horizontal black dotted lines are added for readability only.}
\label{fig:nodes}
\end{figure}

As we mention above, the sense of asymmetry created by turbulent eddies below the upper turning point depends on their position with respect to the nodes and antinodes of the modes. Since the wavelength of the modes is much larger than the spatial extent of the driving region, it is sufficient to study the position of the super-adiabatic peak with respect to the nodes and antinodes of the mode eigenfunction. We illustrate this in Fig. \ref{fig:nodes}, which shows the position of the node and antinode of the radial modes obtained through our model, with respect to the super-adiabatic peak, where the stochastic excitation is mainly located. Below $\nu \sim 3.5$ mHz, the super-adiabatic peak lies above the closest antinode (dashed blue line in Fig. \ref{fig:nodes}), thus generating negative asymmetries. Above the aforementioned threshold, the closest antinode is above the super-adiabatic peak and it generates positive asymmetry. Close to $3.5$ mHz, the super-adiabatic peak coincides with an antinode, so that the asymmetry is very low.

The dichotomy between turbulent eddies below and above the upper turning point stands thus: excitation localised below the upper turning point makes modes with $\nu \lesssim \nu_\text{max}$ negatively asymmetric and modes with $\nu \gtrsim \nu_\text{max}$ positively asymmetric; excitation localised above the upper turning point makes all modes negatively asymmetric; since the source of excitation has a certain spatial extent, the total asymmetry is a combination of both cases. This is in perfect accordance with the dashed blue curve of Fig. \ref{fig:asym_vs_rats} which shows the asymmetry profile $B(\nu)$ with $R_k = \lambda = 1$ (i.e. imposing the same turbulent spectrum everywhere). Indeed, this curve shows that $B$ is negative for low frequencies, and positive for high frequencies. If the source of excitation was only located below the upper turning point of the modes, the threshold between the two regimes would be $\sim 3.5$ mHz; however, turbulent eddies above the upper turning point generate additional negative asymmetry, thus shifting the curve downwards and increasing the threshold between the $B < 0$ and the $B > 0$ regimes.

The results presented in Figs. \ref{fig:asym_vs_k0abs} and \ref{fig:asym_vs_rats} are also easily interpreted. Indeed, Eqs. \eqref{eq:dominant} and \eqref{eq:crossed} show that the efficiency of stochastic excitation scales as $k_0^{-4}$, where $k_0$ is the injection wavenumber of turbulent kinetic energy. Therefore, keeping $R_k$ constant does not change the contribution of atmospheric turbulence relatively to the contribution of turbulence below the upper turning point, and consequently only impacts the mode amplitude, not its asymmetry.

However, decreasing $k_{0,\text{atm}}$ with respect to $k_{0,\text{int}}$ increases the contribution of atmospheric eddies relatively to eddies below the upper turning point. Therefore, increasing $R_k$ makes asymmetries at high frequencies decrease, and the frequency above which $B > 0$ increases. The ratio $R_k$ needed for the asymmetry profile to be negative throughout the entire spectrum is only $R_k \sim 2$, which is explained by the high sensitivity of the excitation efficiency on $k_0$ (since it scales to $k_0^{-4}$).

The physical interpretation of the influence of $\lambda$ on the asymmetry profile $B(\nu)$ is not as straightforward. It cannot be interpreted in the same way as the influence of $k_0(r)$, since we consider $\lambda$ to be uniform throughout the region of excitation. Furthermore, it cannot be interpreted in terms of the relative contribution of the leading and cross term (see Eq. \ref{eq:power}), because both terms scale to $\lambda$. The existence of the threshold $\lambda_\text{sat}$ can be explained as follows: depending on $\lambda$, the typical period of solar $p$-modes compares differently to the typical turbulent eddy-time correlation, that is, $\omega$ compares differently to $\omega_k$, which indeed depends on $\lambda$. Depending on whether $\omega < \omega_k$ or $\omega > \omega_k$, the temporal turbulent spectrum $\chi_k(\omega)$ vary differently with $\omega$: indeed, $\chi_k$ is almost flat for very low frequencies, whereas it decreases as $\omega^{-2}$ for high frequencies. As such, as $\lambda$ is increased, it is expected that the qualitative behaviour of the mode properties - including mode asymmetry - changes when $\omega \sim \omega_k$. This explanation is supported by the value found for $\lambda_\text{sat}$. Indeed, if we take $k \sim 10^{-6}$ m$^{-1}$, $u_k \sim 10^3$ m.s$^{-1}$ and $\omega \sim 10^{-3}$ rad.s$^{-1}$, we obtain $\omega / \omega_k \sim \lambda$. Consequently, $\lambda \sim 1$ does correspond to the threshold at which $\omega$ and $\omega_k$ have the same order of magnitude.

\subsection{Contribution of correlated turbulent perturbations to $B(\nu)$}

Earlier in this paper, we illustrate how a certain degree of correlation between the oscillating modes and the turbulent fluctuations can create mode asymmetry. Furthermore, it has been claimed \citep{nigamCORREL98} that correlation between pulsational velocity and acoustic turbulent perturbations is at the root of the inversion of the sign of mode asymmetry between spectrometric and photometric measurements. This suggests that this correlation plays a crucial role when it comes to interpreting photometric data, although other explanations exist \citep{duvall93, georgobiani03}. However, determining whether or not this role can be disregarded in the velocity spectrum or if it must be taken into account at all (even if it is not so significant as to change the sign of the mode asymmetries) remains an open question.

Our model allows us to shed some light upon this issue. In the following, we compute the asymmetry profile $B(\nu)$ alternatively with and without the cross term $C(\omega)$ in Eq. \eqref{eq:power}. In terms of amplitude, the leading term will, unsurprisingly, dominate over the correlation term $C(\omega)$; however, one must keep in mind that the asymmetry of the mode line profile is a subtle effect, and it is possible that, albeit negligible in amplitude, $C(\omega)$ impacts the asymmetry as much as the leading term.

\begin{figure}
\centering
\includegraphics[width=\linewidth, trim=20 0 0 0]{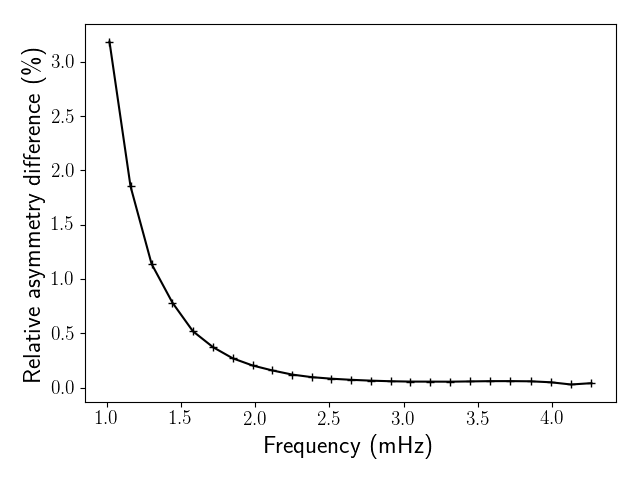}
\caption{Relative difference (in percentages) between the asymmetry parameter $B$ obtained by respectively taking into account and discarding the cross term $C(\omega)$ ($k_{0,\text{int}} = 2 \text{ Mm}^{-1}$, $R_k = 2$ and $\lambda = 1$), as a function of frequency.}
\label{fig:with_without}
\end{figure}

Fig. \ref{fig:with_without} shows the relative difference between the asymmetry profile $B(\nu)$ when the cross term is respectively taken into account and discarded. This relative difference is highest at low frequency, and almost vanishes when $\nu > \nu_\text{max}$. This can be easily explained by the fact that low-frequency modes have the smallest power spectral density, while power spectral density associated to the noise is higher: in contrast, the effect of correlated turbulent perturbations is at its most substantial at the lowest frequencies. Nevertheless, even in the lowest part of the spectrum, the relative difference in asymmetry between a model with and without the cross term does not exceed $3 \%$, which is much smaller than the dispersion characterising observed asymmetries. We therefore conclude that the dominant source of asymmetry in velocity data is the source localisation and that the effect of correlated turbulent perturbations can be disregarded.

It is possible to support this conclusion with a simple order of magnitude estimation. Indeed, $B$ is of order $10^{-2}$, which means it is necessary for the cross term to represent at least $1 \%$ of the leading term to have a significant impact on the asymmetry profile. The Cauchy-Schwarz inequality provides an upper bound to the cross term: $C(\omega) \leqslant \widehat{\vv}_\text{rms}(\omega)\widehat{u}_\text{rms}(\omega)$, where the case of equality happens when the correlation between the mode $\vv$ and the turbulence $u$ is optimal. Therefore

\begin{equation}
\dfrac{\left\langle \left|\widehat{\vv}(\omega)\right|^2\right\rangle}{C(\omega)} \gtrsim \dfrac{\widehat{\vv}_\text{rms}(\omega)}{\widehat{u}_\text{rms}(\omega)}~,
\end{equation}
where $C(\omega)$ corresponds to the second term in Eq. \eqref{eq:power}, and is defined by Eq. \eqref{eq:crossed}.

If we estimate the power spectral density of the mode as $\widehat{\vv}_\text{rms}^2 \sim 10^5~\text{m}^2\text{.s}^{-2}\text{.Hz}^{-1}$ and that of the turbulent fluctuations as $\widehat{u}_\text{rms}^2 \sim 10~\text{m}^2\text{.s}^{-2}\text{.Hz}^{-1}$ \citep[fig.~2]{turck04}, the above ratio becomes $\sqrt{10^5 / 10} \sim 10^2$. We note that at this stage, the correlation term is about $1 \%$ of the leading term, which is barely enough to impact the asymmetries significantly. However, by considering the limiting case in the Cauchy-Schwarz inequality, we assumed that the mode velocity and the turbulent velocity were both optimally coherent and completely independent. This is not, however, the case; consequently we probably overestimated the importance of $C(\omega)$ by at least an order of magnitude. Therefore, this crude order of magnitude estimation indeed tends to support the conclusion that correlated turbulent perturbations can be disregarded when interpreting mode asymmetries in the velocity spectrum.

\section{Comparison with observations}\label{sec:observations}

Observed properties of solar-like oscillations depend not only on the observable (velocity or intensity), but also on the specifics of each instruments. As far as velocity measurements are concerned, all instruments do not perform their Doppler observations on the same spectral line. Since different spectral lines form at different altitudes in the atmosphere, and since the properties of turbulence change throughout the atmosphere, mode properties, and especially mode asymmetries, may depend on the instrument.

In the following, we compare the results of our model to observations performed with the \textit{GONG} network. We then focus on the dependence of the asymmetry profile $B(\nu)$ on the height at which the velocity spectrum is observed. We use the the `numerical spectrum' model (see Sect. \ref{sec:methods}) to compare the asymmetry profiles $B(\nu)$ as observed by several instruments performing solar velocity spectrum measurements. Finally, we focus on the bias in the determination of mode eigenfrequencies entailed by mode asymmetry, whose understanding is of primary importance for accurate inference of mode properties.

\subsection{Comparison with \textit{GONG} observations}

First - and in order to support the validity of our model in terms of mode asymmetry - we compare the asymmetries yielded by our model to those inferred from observations. We use the data points extracted from the spectrum analysis of \citet{barban04}. The authors chose an equivalent, albeit different, set of parameters to fit the acoustic modes observed by the \textit{GONG} network. Therefore, we reconstructed the shape of the modes point by point using the parameters extracted from their fit and fitted them again with the formula given by Eq. \eqref{eq:formula}. We note that the modes analysed in their study are not radial, but have angular degrees ranging from $l=15$ to $l=50$. It is known, however, that the dependence of mode asymmetry on the angular degree is very weak \citep[see for instance][who studied modes of angular degree up to $l = 170$]{duvall93} because the eigenfunctions associated to the acoustic modes are very weakly dependent on $l$ close to the photosphere, as long as $l$ is not too high. Thus, comparing our radial study to their non-radial observations remains relevant.

\begin{figure}
\includegraphics[width=\linewidth, trim=20 0 0 0]{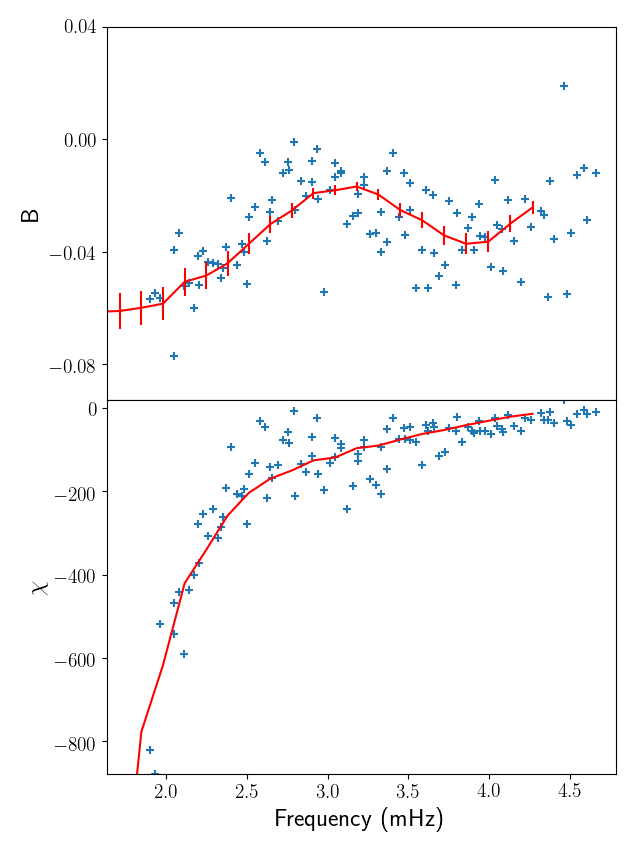}
\caption{Asymmetry profile obtained by the `numerical spectrum' model (\textit{solid red line}), compared to the observed asymmetry profile (\textit{blue crosses}). The data points are taken from \citet{barban04}. For more readability, only data points corresponding to modes with angular degrees $15 \leqslant l \leqslant 20$ have been retained. The asymmetry profile is given in terms of the parameter $B$ (\textit{top panel}), and alternatively in terms of the parameter $\chi$ (\textit{bottom panel}), which is defined by Eq. \eqref{eq:chi}. The error bars in the top panel correspond to the uncertainty on the observed values of the mode linewidths, which propagates to the asymmetry parameter $B$.}
\label{fig:us_vs_caro}
\end{figure}

We showcase this comparison in Fig. \ref{fig:us_vs_caro}. The parameter $\lambda$ was adjusted for a better agreement with the observations. We obtained $\lambda = 0.5$ (see Eq. \ref{eq:lambda} for a definition of this parameter), which is approximately the value obtained by constraining mode amplitudes \citep{samadiG01}. It is rather clear that we reproduce the main features of the asymmetry profile $B(\nu)$, especially its sense of variation on the different intervals (when the frequency range of the observations overlaps ours), as well as the positions and values of its local extrema. Alternatively, we show the same comparison in terms of the asymmetry parameter $\chi$ in the bottom panel of Fig. \ref{fig:us_vs_caro}. This parameter is defined by Eq. \eqref{eq:chi}, and is more robust when it comes to comparing theory and observations because it does not depend on the determination of the mode linewidths, which may introduce extra uncertainty in the determination of the observed asymmetries. In conclusion, the good agreement obtained between our model and observations show that the model developed in this paper is relevant to account for acoustic mode asymmetry quantitatively.

\subsection{Dependence of asymmetry profile $B(\nu)$ on observation height}

We consider three different observation heights in this paper associated, respectively, with the \textit{Michelson Doppler Imager} (MDI), the \textit{GOLF} instrument (both onboard the \textit{SOHO} spacecraft), the \textit{Global Oscillation Network Group} (ground-based), and the \textit{Helioseismic and Magnetic Imager} (HMI; onboard the \textit{SDO} spacecraft). We provide further information on the observation height of each of these instruments in Table \ref{table:heights}. However, determining the formation height of a given absorption line is extremely difficult in that it does not depend only on the nature of the line \citep{fleck11}. Therefore, the values given in Table \ref{table:heights} are not to be considered as precise estimates but, rather, as approximate figures.

\begin{table}
\centering
\begin{tabular}{ccccc}
Instrument & Line & $\lambda$ ($\AA$) & Height (km) \\
\hline\hline \\
MDI/GONG & Ni I & 6768 & 170 \\
GOLF & Na I D1/D2 & 5896/5890 & 340 \\
HMI & Fe I & 6173 & 100 \\
\end{tabular}
\caption{Summary of the nature, wavelength, and formation height of the absorption line used by the instruments considered in this paper. The formation heights are given with respect to the photosphere. References: \citet{fleck11} (MDI/GONG, HMI) and \citet{baudin05} (GOLF). Note that the \textit{MDI} instrument and the \textit{GONG} network use the same spectral line and therefore observe the modes at the same altitude.}
\label{table:heights}
\end{table}

\begin{figure}
\includegraphics[width=\linewidth, trim=20 0 0 0]{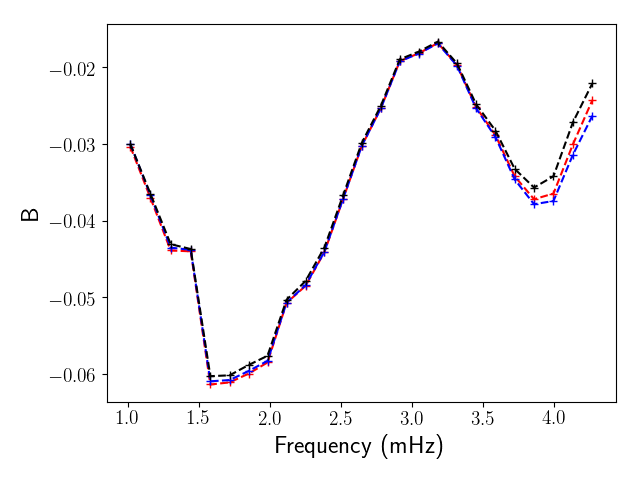}
\caption{Asymmetry profile $B(\nu)$ obtained by the `numerical spectrum' model (with $\lambda = 0.5$), as would be observed at three different heights in the atmosphere, corresponding respectively to the observation heights of \textit{GOLF} (\textit{dashed red line}), \textit{MDI}/\textit{GONG} (\textit{dashed blue line}) and \textit{HMI} (\textit{dashed black line}).}
\label{fig:compare_inst}
\end{figure}

In Fig. \ref{fig:compare_inst}, we compare the asymmetry profile $B(\nu) $ as it would be observed by the various instruments listed in Table \ref{table:heights}. We control the height by tuning the radial coordinate $r_\text{o}$ at which the Green's function of the homogeneous wave equation is calculated (see Sect. \ref{sec:methods} for more details). Mode asymmetry is only slightly dependent on the observation height. This is not so surprising since mode asymmetry is a global property of the modes. It is noticeable, however, that the difference between these asymmetry profiles is most prominent at high frequency ($\gtrsim 3.5$ mHz). This is because by observing the modes at a lower height in the atmosphere, we effectively increase the contribution of atmospheric turbulence compared to the contribution of sub-photospheric turbulence on mode asymmetry. As we discussed in Sect. \ref{sec:discussion}, it is at high frequency that the atmospheric turbulence has a more significant impact on mode asymmetry. Therefore, changing the observation height has more impact at high frequency than at low frequency.

\subsection{Bias in eigenfrequency determination\label{sec:bias}}

In order to infer the stellar internal structure from asteroseismic measurements, it is important to determine the eigenfrequencies of the observed modes not only precisely, but also accurately. As was noted early on \citep{duvall93, abramsK96, chaplin99, thiery00, toutain98}, using a symmetric, Lorentzian model to fit asymmetric line profiles results in an appreciable bias in the best-fit parameters, thus rendering inaccurate the frequency determination.

We illustrate this bias in the following. We fit the line profiles we obtained numerically with a formula given either by Eq. \eqref{eq:formula} or by the symmetric version (i.e. fixing $B=0$). Following \citet{abramsK96}, we denote the reduced frequency bias as

\begin{equation}
\delta \nu = \dfrac{\nu_0^{B=0} - \nu_0^{B \neq 0}}{\Gamma_{\nu_0}}~.
\label{eq:bias}
\end{equation}

\begin{figure}
\includegraphics[width=\linewidth, trim=20 0 0 0]{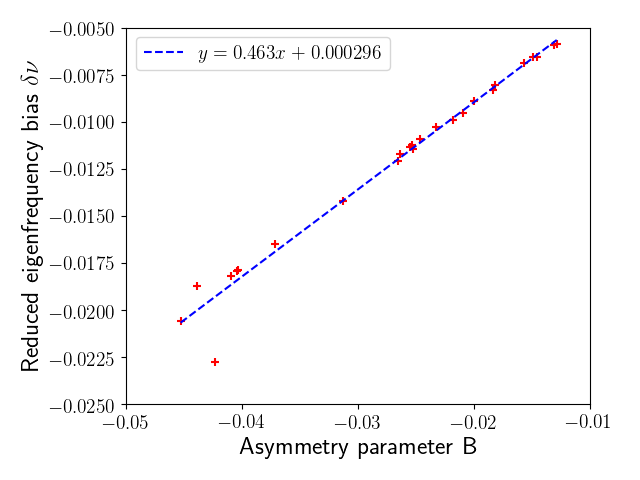}
\caption{Reduced frequency bias $\delta\nu$, defined by Eq. \eqref{eq:bias}, as a function of the asymmetry parameter $B$ for each radial $p$-mode between $n=6$ and $n=30$. The blue dashed line shows the best linear fit.}
\label{fig:bias}
\end{figure}

Figure \ref{fig:bias} shows the results obtained for all radial modes described in Sect. \ref{sec:results}. Given that the asymmetry of solar $p$-modes in the velocity spectrum are all negative, using a symmetric fit introduces an underestimation of $\nu_0$. It is very clear that this underestimation grows linearly with the asymmetry $B$ of the modes. We superimpose on Fig. \ref{fig:bias} the best linear fit to the numerical data, given by: $\delta\nu = 0.463~B + 0.000296$. Given the typical values of $\delta\nu$, it is safe to say that the intercept of this linear regression is negligible.

This result can be easily interpreted by taking a closer look to the asymmetric fitting formula given by Eq. \eqref{eq:formula}. Indeed, after applying elementary algebra, it becomes clear that for leading order in $B$, this expression reaches its maximum at $x \sim B$ (or $\nu_\text{max} = \nu_0~+~B\Gamma_{\nu_0} / 2$). Therefore, while the asymmetric fit accurately finds the eigenfrequency at $\nu_0^{B \neq 0} = \nu_0$, the symmetric fit finds it at $\nu_0^{B=0} = \nu_\text{max}$. One can therefore derive the simple expression $\delta\nu = B/2$, which is in accordance with the linear fit shown in Fig. \ref{fig:bias}.

The frequency bias introduced when not taking the line profile asymmetry into account can easily reach several percents of the mode linewidth. This bias largely exceeds the frequency resolution achievable in current solar measurements, especially for high-frequency modes, which are the widest.

\section{Summary and conclusion}\label{sec:conclusion}

In this paper, we detail the development of a realistic and predictive model for the asymmetry displayed by solar radial $p$-modes in the velocity spectrum. The basic idea behind this model is to compute the Green's function associated to the radial acoustic wave equation, as well as its inhomogeneous part (which corresponds to the source of excitation) and to convolve the two to reconstruct the velocity power spectral density point by point. Once the power spectral density is reconstructed, we extract its resonant modes and study their asymmetry. In particular, and unlike previous attempts to such modelling, we included in our model the correlation of the oscillating modes with the fluctuations associated to turbulent velocity.

First, the Green's function associated with the wave equation was computed numerically. We put the wave equation in the form of a 1D stationary Schrödinger equation, whose potential only depends on the equilibrium structure of the Sun. We extracted the acoustic potential from a solar patched model: the solar interior is calculated using the 1D evolutionary code CESTAM and the solar atmosphere is calculated using the 3D hydrodynamic code CO$^5$BOLD and horizontally averaged. We integrated the wave equation along the solar radius, and added a point-like, normalised source to the integration scheme in order to compute the Green's function.

Secondly, the source term of the wave equation being of stochastic nature, we modelled the statistical properties of the source by means of theoretical developments. We made use of the adequate closure relation to express the third and fourth-order correlation products of the turbulent velocity as functions of second-order products; more specifically on their spatial and temporal Fourier transform. We developed two distinct models: one is based on theoretical prescriptions for the spatial and temporal spectrum of turbulent kinetic energy; the other is based on theoretical modelling of the temporal spectrum only, whereas the spatial spectrum is extracted from a 3D hydrodynamic simulation of the solar atmosphere. We refer to the former as the `theoretical spectrum' model, and to the latter as the `numerical spectrum' model.

The asymmetry $B$ displayed by the modes in our model drastically depends on their frequency $\nu$. This is because the shape of the eigenfunctions close to the photosphere is very dependent on $\nu$. We find that $B$ is negative throughout the $p$-mode spectrum, and that its behaviour weakly depends on the input parameters of our model. It drops from $-0.01$ to $-0.05$ between $1$ mHz and $1.7$ mHz, then rises to $0.015$ at $3$ mHz, and decreases again from $3$ mHz to $4$ mHz. Above $4$ mHz, the behaviour of $B(\nu)$ is much more dependent on the value given to our input parameters and, in particular, on the injection scale associated to the turbulent cascade above the photosphere, compared to below the photosphere. This is related to the fact that the contribution of atmospheric turbulence to mode excitation only becomes significant at high frequency, so that only in this part of the spectrum it may have an impact on mode asymmetry.

The asymmetry of the modes can have two different origins: localisation of their source of excitation within a region of lesser spatial extent than the mode wavelength and correlation between the oscillating modes and the fluctuations associated to turbulent velocity. Formally, these two phenomena have the same impact on mode asymmetry, so that they cannot be separated using observational data only. Our model allows us to make this distinction and to study their relative weight in the total mode asymmetry. We find that the correlation with turbulent fluctuations is negligible in the velocity spectrum, and that the observed asymmetries are exclusively due to source localisation. More precisely, we interpret the results of our model in terms of the source position with respect to the various nodes and antinodes featured by the eigenfunctions of the modes. In the case of a point-like source of excitation, mode asymmetry drastically depends on whether it is located within or outside the mode acoustic cavity. In our model, however, the source of excitation has a certain spatial extent, so that the total asymmetry is a combination of the contributions from the source outside and inside the mode acoustic cavity.

We find that it is impossible to interpret even the qualitative behaviour of the asymmetry profile $B(\nu)$ by considering that the source of excitation is point-like (either outside or inside the modes cavity). On the contrary, taking into account the spatial extent of the source allows us to reproduce the observed asymmetries, not only qualitatively, but also quantitatively. This positive result shows that our model is indeed relevant to describe - and, more importantly, to predict - acoustic mode asymmetry in solar-like oscillators. It also shows that any model that assumes a point-like source of excitation cannot give reliable results as far as mode asymmetry is concerned. In particular, such a model would predict positive asymmetries for high-frequency modes, whereas observations show that all asymmetries are negative when measured in terms of velocity power spectral density.

Finally, we study the eigenfrequency bias entailed by neglecting to fit observations with an asymmetric profile. We find that for the most asymmetric modes, this bias can reach several percent of the mode linewidth. Therefore, this bias is higher for high frequency modes, which are the widest. In particular, for $\nu \sim 4$ mHz, the asymmetry parameter is of order $B \sim -0.04$, and the linewidth is of order $\Gamma \sim 10~\mu$Hz, so that the eigenfrequency bias is of order $\delta\nu \sim 0.2~\mu$Hz. This is in perfect accordance with actual biases obtained from observation fit of the solar spectrum \citep[see][Fig. 6, topmost panel]{benomar18}. Since the eigenfrequency bias is most pronounced for higher frequency (because it is proportional to the mode line-width, which is widest at high frequency), it is likely to have a non-negligible impact on inversion methods, especially those based on asymptotic formulae. One must keep in mind, however, that the deviation of the modelled eigenfrequencies from the observed ones, induced by surface effects, largely dominates the eigenfrequency bias entailed by symmetric fits.

In this paper, we have restricted ourselves to the study of solar radial $p$-modes. Our formalism can be easily adapted to the study of non-radial modes simply by using a non-radial wave equation instead of the radial one. However, since the eigenfunction associated to $p$-modes in solar-like oscillators are very weakly independent on angular degree $l$ close to the photosphere, which is precisely where the excitation takes place, the mode asymmetry is not expected to vary significantly with $l$, at least as long as $l$ remains reasonably small. Observational data tend to confirm this \citep[see e.g. ][who report that the spectral parameters of individual modes collapse to slowly varying functions of frequency only for modes with $l \lesssim 100 $]{vorontsovJ13}.

We only considered one type of acoustic source in this study, that is, the turbulent fluctuations of the Reynolds stress. Indeed, it has been shown by \citet{steinN01} that this is the dominant source of excitation of solar acoustic modes \citep[see also][]{chaplin05, samadi07, nordlund09}. Therefore, our objective was to start by considering only this source. However, further refinements of the model will have to include other sources of excitation, in the form of non-adiabatic, turbulent pressure fluctuations.

Our formalism can also be easily applied to other solar-like oscillators. Comparing the asymmetries featured by the velocity spectra of several solar-like oscillators as modelled by the method presented in this paper and, in particular, the trend followed by mode asymmetry with stellar parameters such as effective temperature or surface gravity, undoubtedly constitutes the next step of this study. In the long run, mode asymmetry may serve as a useful tool for seismic diagnoses of solar-like oscillators. However, the one major difference that remains between the solar case and other stars is that the Sun is the only solar-like oscillator for which spectra obtained by spectrometric measurements are sufficiently resolved to allow for a determination of their mode asymmetry. The asymmetry of acoustic modes of all other stars can only be observed in intensity spectra. As has been reported numerous times \citep[see e.g. ][]{duvall93}, asymmetry in intensity and in velocity spectra are drastically different. It is, therefore, necessary to adapt our formalism to the intensity spectrum, which is another key element of any further considerations on the matter treated here.

\begin{acknowledgements}
J.P wishes to warmly thank Louis Manchon for having provided us with the solar patched model we used in this paper. J.P and K.B would also like to thank Marie-Jo Goupil for her thorough reading of the manuscript, as well as John Leibacher for useful discussions. J.P and K.B acknowledge financial support from the `Programme National de Physique Stellaire' (PNPS) of CNRS/INSU.
H.G.L. acknowledges financial support by the Sonderforschungsbereich SFB\,881 `The Milky Way System' (subprojects A4) of the German Research Foundation (DFG).
\end{acknowledgements}

\bibliographystyle{aa}
\bibliography{biblio}

\begin{appendix}

\section{The inhomogeneous wave equation (Eq. \ref{eq:wave_equation})}\label{app:wave_eq}

\subsection{Hydrodynamic equations and their linearisation}

We linearise the governing, hydrodynamic equations in order to derive the wave equation with its source term. We consider that the mode velocity and the turbulent velocity obey separately their own continuity equation. Furthermore, we only consider radial modes, such that the mode velocity may we written in terms of the radial fluid displacement as $\bm{\vv_\text{osc}} = \diff\xi_r / \diff t~\bm{e_r}$. In this context, the governing equations are as follows:

\begin{itemize}

\item[$\bullet$] the continuity equation associated to the mode velocity can be written as
\begin{equation}
\dfrac{\partial \rho}{\partial t} + \bm{\nabla}(\rho\bm{\vv_\text{osc}}) = 0~.
\end{equation}
Writing $\rho = \rho_0 + \rho'$ (where $\rho'$ is the Eulerian density perturbation corresponding to the mode), linearising this equation around a motionless state, and integrating with respect to time yields
\begin{equation}
\rho' + \bm{\nabla}(\rho_0\bm{\xi}) = 0~.
\end{equation}
We then introduce the Lagrangian density perturbation $\delta\rho = \rho' + (\bm{\xi}.\bm{\nabla})\rho_0$, which allows us to write the linearised continuity equation in its final form:
\begin{equation}
\dfrac{\delta\rho}{\rho_0} + \dfrac{1}{r^2}\dfrac{\diff (r^2\xi_r)}{\diff r} = 0~;
\end{equation}

\item[$\bullet$] the Euler equation:
\begin{equation}
\dfrac{\partial \rho\bm{\vv}}{\partial t} + \bm{\nabla}:(\rho\bm{\vv}\bm{\vv}) = -\bm{\nabla}P + \rho\bm{g}~.
\end{equation}
Unlike what we did for the continuity equation, the velocity $\bm{\vv}$ now includes the mode velocity $\bm{\vv_\text{osc}}$ as well as the turbulent velocity $\bm{u_\text{turb}}$. We further decompose the latter into a mean value $\bm{U} \equiv \langle \bm{u_\text{turb}} \rangle$ (where the notation $\langle . \rangle$ refers to an ensemble average) and fluctuations around this mean value $\bm{u} \equiv \bm{u_\text{turb}} - \bm{U}$. As such, we have
\begin{equation}
\bm{\vv} = \bm{U} + \bm{\vv_\text{osc}} + \bm{u} = \bm{U} + \diff\xi_r / \diff t~\bm{e_r} + \bm{u}~.
\end{equation}
The last two terms are treated as small perturbations compared to the first one. In the term $\partial \rho\bm{\vv} / \partial t$, the contribution of $\bm{U}$ vanishes because we consider that $\bm{U}$ is independent of time (in other words, we consider a stationary turbulence), and the contribution of $\bm{u}$ vanishes after ensemble averaging. Concerning the advection term, among the 9 terms of its development, only 2 survive after the linearisation and ensemble averaging, namely $\bm{\nabla}:(\rho\bm{U}\bm{U})$ and $\bm{\nabla}:(\rho\bm{u}\bm{u})$. The first one can be rewritten as $\bm{\nabla} p_t$, where $p_t$ is the turbulent pressure, and is of order zero, so that it will only impact the equilibrium structure. The second one can be equivalently rewritten as $\bm{\nabla} p_t'$, where $p_t'$ refers to the perturbation of the turbulent pressure. Finally, performing a Fourier transform with respect to time, the radial component of the Euler equation reads:
\begin{equation}
-\omega^2\xi_r + \dfrac{1}{\rho_0}\dfrac{\diff p'}{\diff r} + \dfrac{\rho'}{\rho_0}g_0 - g' = -\dfrac{1}{\rho_0}\dfrac{\diff p_t'}{\diff r}~,
\end{equation}
where $p'$ is the Eulerian pressure perturbation, $g_0$ is the mean gravitational acceleration, $g'$ is its Eulerian perturbation and $\diff p_t' / \diff r$ refers to the turbulent fluctuations of the Reynolds stress around its mean value. Since we only aim at modelling radial modes, using the Cowling approximation to eliminate $g'$ would not reduce the order of the final wave equation, and is therefore of no particular use. Instead, we follow \citet{unno89} and express $g'$ as a function of the radial fluid displacement (see their Eq. 14.36):
\begin{equation}
g' = -\dfrac{\diff \phi'}{\diff r} = 4\pi G\rho_0\xi_r~.
\end{equation}
One can note that this is equivalent to saying that the Lagrangian perturbation of the gravitational potential is zero.

\item[$\bullet$] the equation of state we will use to close the system: after some algebra, a linearised version of the equation of state in terms of the Lagrangian perturbations can be derived:
\begin{equation}
\dfrac{\delta \rho}{\rho_0} = \dfrac{1}{\Gamma_1}\dfrac{\delta p}{p_0} - \dfrac{\rho_0 T_0}{p_0}\nabla_\text{ad}\delta s~,
\end{equation}
where $\delta s$ corresponds to the turbulent fluctuation of the specific entropy of the fluid and we define the various thermodynamic coefficients as
\begin{equation}
\begin{array}{ll}
\Gamma_1 &\equiv \left(\dfrac{\partial \ln p}{\partial \ln \rho}\right)_s \\
\nabla_\text{ad} &\equiv \left(\dfrac{\partial \ln T}{\partial \ln p}\right)_s .
\end{array}
\end{equation}
In order to facilitate the following calculations, we replace the Lagrangian pressure perturbation $\delta p$ with the Eulerian one $p'$, and we derive two versions of the linearised equation of state, one with the Lagrangian density perturbation, one with the Eulerian one. Noting that the hydrostatic equilibrium gives us
\begin{equation}
\dfrac{\diff p_0}{\diff r} = -\rho_0 g_0~,
\end{equation}
and that by definition of the Brunt-Väisälä frequency, we have
\begin{equation}
\dfrac{N^2}{g_0} = \dfrac{1}{\Gamma_1}\dfrac{\diff \ln p_0}{\diff r} - \dfrac{\diff\ln \rho_0}{\diff r}~,
\end{equation}
and we finally obtain
\begin{equation}
\begin{array}{ll}
\dfrac{\delta\rho}{\rho_0} &= \dfrac{1}{\Gamma_1}\dfrac{p'}{p_0} - \dfrac{g_0\rho_0}{\Gamma_1 p_0}\xi_r - \dfrac{\rho_0 T_0}{p_0}\nabla_\text{ad} \delta s \\
\dfrac{\rho'}{\rho_0} &= \dfrac{1}{\Gamma_1}\dfrac{p'}{p_0} + \dfrac{N^2}{g_0}\xi_r - \dfrac{\rho_0 T_0}{p_0}\nabla_\text{ad} \delta s~.
\end{array}
\label{eq:eos}
\end{equation}

\end{itemize}

\subsection{Changing variables}

The two variables that we wish to keep in these equations are $\xi_r$ and $p'$. We first make use of Eq. \eqref{eq:eos} to eliminate the density fluctuations. Noting that $c^2 = \Gamma_1 p_0 / \rho_0$ (where $c$ is the sound speed), the continuity and Euler equations then yield:

\begin{equation}
\begin{array}{l}
\dfrac{\diff (r^2\xi_r)}{\diff r} - g_0\dfrac{r^2}{c^2}\xi_r + \dfrac{r^2}{\rho_0 c^2}p' = r^2\nabla_\text{ad}\dfrac{\rho_0 T_0}{p_0}\delta s \\
\dfrac{1}{\rho_0}\dfrac{\diff p'}{\diff r} + \dfrac{g_0}{c^2}\dfrac{p'}{\rho_0} + (N^2 - \omega^2 -4\pi G\rho_0)\xi_r = \dfrac{\rho_0 g_0 T_0}{p_0}\nabla_\text{ad} \delta s - \dfrac{1}{\rho_0}\dfrac{\diff p_t'}{\diff r}~.
\end{array}
\label{eq:density_discarded}
\end{equation}

In order to remove $\xi_r$ from the 0-th order term in the $\xi_r$ equation, and same for $p'$, the required variable change is then \citep{unno89}:

\begin{equation}
\begin{array}{l}
r^2\xi_r(r) = \widetilde{\xi}(r) \exp\left(\displaystyle\int_0^r \dfrac{g_0}{c^2}~\diff r'\right) \\
p' = \rho_0 \widetilde{\eta}(r) \exp\left(\displaystyle\int_0^r \dfrac{N^2}{g_0}~\diff r'\right) .
\end{array}
\end{equation}

Plugging this into Eq. \eqref{eq:density_discarded}, we obtain

\begin{multline}
\dfrac{\diff\widetilde{\xi}}{\diff r} + \dfrac{r^2}{c^2}\exp\left(\displaystyle\int_0^r \dfrac{N^2}{g_0}-\dfrac{g_0}{c^2}~\diff r'\right)\widetilde{\eta} \\
= r^2 \exp\left(-\displaystyle\int_0^r \dfrac{g_0}{c^2}~\diff r'\right) \nabla_\text{ad}\dfrac{\rho_0 T_0}{p_0} \delta s~,
\label{eq:plugging1}
\end{multline}
and

\begin{multline}
\dfrac{\diff \widetilde{\eta}}{\diff r} + \dfrac{1}{r^2}\exp\left(\displaystyle\int_0^r \dfrac{g_0}{c^2}-\dfrac{N^2}{g_0}~\diff r'\right) (N^2-\omega^2 -4\pi G\rho_0)\widetilde{\xi} \\
= \exp\left(-\displaystyle\int_0^r \dfrac{N^2}{g_0}~\diff r'\right) \left[\nabla_\text{ad}\dfrac{\rho_0 g_0 T_0}{p_0} \delta s - \dfrac{1}{\rho_0}\dfrac{\diff p_t'}{\diff r}\right]~.
\label{eq:plugging2}
\end{multline}
where we denote the right-hand side terms of Eqs. \eqref{eq:plugging1} and \eqref{eq:plugging2} as $S_0$ and $S_1$ respectively in the following. We also define
\begin{equation}
\begin{array}{ll}
I(r) &\equiv \exp\left(\displaystyle\int_0^r \dfrac{N^2}{g_0}-\dfrac{g_0}{c^2}~\diff r'\right) \\
x(r) &\equiv \dfrac{r\sqrt{I}}{c} \\
k^2 &\equiv \dfrac{\omega^2 - N^2 + 4\pi G\rho_0}{c^2}~.
\end{array}
\end{equation}
The above set of equations can be rewritten as
\begin{equation}
\begin{array}{l}
\dfrac{\diff\widetilde{\xi}}{\diff r} + x^2\widetilde{\eta} = S_0 \\
\dfrac{\diff\widetilde{\eta}}{\diff r} - \dfrac{k^2}{x^2}\widetilde{\xi} = S_1~.
\end{array}
\label{eq:new_coupled}
\end{equation}

We can now eliminate $\widetilde{\eta}$ to get a single second-order wave equation. Using the first of Eq. \eqref{eq:new_coupled} to express $\widetilde{\eta}$ as a function of $\widetilde{\xi}$, and plugging it in the second equation, we get the following equation:

\begin{equation}
\dfrac{\diff^2\widetilde{\xi}}{\diff r^2} - \dfrac{2}{x}\dfrac{\diff x}{\diff r}\dfrac{\diff\widetilde{\xi}}{\diff r} + k^2\widetilde{\xi} = \dfrac{\diff S_0}{\diff r} - \dfrac{2}{x}\dfrac{\diff x}{\diff r}S_0 - x^2S_1~.
\label{eq:second_order_origin}
\end{equation}

Similarly to what has been done for the first change of variables, we wish for the left-hand side to contain no first-order term, but only second-order and 0th-order ones. Thus we introduce yet another variable: $\Psi(r) \equiv \widetilde{\xi} / x$. Plugging this new variable into Eq. \eqref{eq:second_order_origin}, we easily obtain a wave equation that assumes the form of a 1D stationary Schrödinger equation
\begin{equation}
\dfrac{\diff^2\Psi}{\diff r^2} + \left(\dfrac{\omega^2}{c^2} - V(r)\right)\Psi = \dfrac{1}{x}\left(\dfrac{\diff S_0}{\diff r} - \dfrac{2}{x}\dfrac{\diff x}{\diff r}S_0 - x^2S_1\right)~,
\label{eq:schrodinger}
\end{equation}
with an acoustic potential $V(r)$ that only depends on the star's equilibrium state:
\begin{equation}
V(r) = \dfrac{N^2 - 4\pi G\rho_0}{c^2} + \dfrac{2}{x^2}\left(\dfrac{\diff x}{\diff r}\right)^2 - \dfrac{1}{x}\dfrac{\diff^2x}{\diff r^2}~.
\label{eq:potential}
\end{equation}

\subsection{The source term}

With the above notations, the parameters intervening in the source term of Eq. \eqref{eq:schrodinger} have the following expressions:

\begin{equation}
\begin{array}{l}
S_0(r) = r^2 \exp\left(-\displaystyle\int_0^r \dfrac{g_0}{c^2}~\diff r'\right) \nabla_\text{ad}\dfrac{\rho_0 T_0}{p_0}\delta s \\
S_1(r) = \exp\left(-\displaystyle\int_0^r \dfrac{N^2}{g_0}~\diff r'\right) \left[\nabla_\text{ad}\dfrac{\rho_0 g_0 T_0}{p_0} \delta s - \dfrac{1}{\rho_0}\dfrac{\diff p_t'}{\diff r}\right] \\
x(r) = \dfrac{r}{c}\exp\left(\dfrac{1}{2}\displaystyle\int_0^r \dfrac{N^2}{g_0}-\dfrac{g_0}{c^2}~\diff r'\right) .
\end{array}
\end{equation}

Furthermore, one can easily derive the following relationship between $\nabla_\text{ad}$ and $\alpha_s \equiv (\partial P / \partial s)_\rho$ by means of the adequate Schwarz relation:
\begin{equation}
\nabla_\text{ad} \dfrac{\rho_0 T_0}{p_0} = \dfrac{\alpha_s}{\rho_0 c^2}~.
\end{equation}
After some manipulations, one finally obtain the source term in the form:
\begin{multline}
S(r) = \dfrac{r}{c\rho_0} \exp\left(-\dfrac{1}{2}\displaystyle\int_0^r \dfrac{N^2}{g_0}+\dfrac{g_0}{c^2}~\diff r'\right) \\
\times \left[\alpha_s \delta s \dfrac{\diff}{\diff r}\ln\left(\dfrac{\alpha_s \delta s}{\rho_0}\right) - \alpha_s \delta s\left(\dfrac{N^2}{g_0}+\dfrac{g_0}{c^2}\right) + \dfrac{\diff p_t'}{\diff r}\right] .
\end{multline}

Finally, since
\begin{equation}
\dfrac{N^2}{g_0} + \dfrac{g_0}{c^2} = -\dfrac{\diff\ln\rho_0}{\diff r}~,
\end{equation}
this expression can be drastically simplified to
\begin{equation}
S(r) = \dfrac{r}{c\sqrt{\rho_0(r=0)\rho_0(r)}}\left(\dfrac{\diff (\alpha_s \delta s)}{\diff r} + \dfrac{\diff p_t'}{\diff r}\right) .
\label{eq:source}
\end{equation}

This form clearly shows that the source term can be split three ways: a monopolar source term (proportional to $\delta s \diff\alpha_s / \diff r$) due to non-adiabatic pressure fluctuations in a stratified environment, a dipolar term (proportional to $\alpha_s \diff\delta s / \diff r$) due to a stratification in the non-adiabatic pressure fluctuations themselves, and a quadripolar term (proportional to $\diff p_t' / \diff r$) due to Reynolds stress fluctuations. In the following, we only consider this last term but we also show here how the effect of non-adiabaticity can be introduced as well.

To conclude, note that the value of the fluid density at the centre of the star $\rho_0(r=0)$ appears both in the definition of the variable $\Psi$ and in the source term $S(r)$. This is due to the particular change of variable we have performed, and it can be factored out of the wave equation. Finally, we can put the wave equation in the following form:

\begin{equation}
\dfrac{\diff^2\Psi}{\diff r^2} + \left(\dfrac{\omega^2}{c^2} - V(r)\right)\Psi = S(r)~,
\label{eq:wave_eq_app}
\end{equation}
with $V(r)$ given by Eq. \eqref{eq:potential}, and the source term and wave variable are given by

\begin{equation}
\begin{array}{l}
S(r) = \dfrac{r}{c\sqrt{\rho_0(r)}}\left(\dfrac{\diff (\alpha_s \delta s)}{\diff r} + \dfrac{\diff p_t'}{\diff r}\right) \\
\Psi(r) = rc(r)\sqrt{\rho_0(r)}\xi_r(r)~.
\end{array}
\end{equation}

\section{From the Green's function to the power spectral density}\label{app:spectrum}

Here we detail the calculations carried out to obtain the expression of the velocity power spectral density (Eq. \ref{eq:power}) as a function of the Green's function associated with the homogeneous wave equation \eqref{eq:wave_equation}. Note that these calculations correspond to the `theoretical spectrum' model described in Section \ref{subsec:leading}. The calculations in the `numerical spectrum' model being fairly similar, we do not detail it. We start with the development given by Eq. \eqref{eq:dev}, with the expression of $\widehat{\vv_\text{osc}}$ given by Eq. \eqref{eq:v}. We detail the treatment of both terms in the development \eqref{eq:dev} (leading term and cross term) separately.

\subsection{The leading term}

For more clarity, in the following, we introduce
\begin{equation}
X_\omega(\bm{r}) \equiv G_\omega(\bm{r})\dfrac{||\bm{r}||}{c(\bm{r})\sqrt{\rho_0(\bm{r})}}~.
\end{equation}
We then have
\begin{multline}
\left\langle \left|\widehat{\vv_\text{osc}}(\omega)\right|^2\right\rangle = \left\langle \dfrac{\omega^2}{r_\text{o}^2 c_\text{o}^2 \rho_0(r_\text{o})}\right. \\
\times\left.\displaystyle\iint \diff^3\bm{r_{s1}}\diff^3\bm{r_{s2}}~\left(\bm{\nabla} X_\omega . \rho_0\widehat{u_r \bm{u}}\right)(\bm{r_{s1}}) \left(\bm{\nabla} X_\omega^{\star} . \rho_0\widehat{u_r \bm{u}}^{\star}\right)(\bm{r_{s2}}) \right\rangle~,
\end{multline}
where $r_\text{o}$ is the radius at which the spectrum is observed, $c_\text{o}$ is the speed of sound at that radius, and the notation $^\star$ refers to the complex conjugate.

We then perform the following change of variable: $\bm{R} = (\bm{r_{s1}} + \bm{r_{s2}})/2$ and $\bm{r} = (\bm{r_{s1}} - \bm{r_{s2}})/2$, the former being a `slow' variable, and the latter a `fast' variable. This allows us to separate the scales relevant to the turbulent velocity $\bm{u}$ from the scales relevant to the medium stratification and the mode wavelength, with turbulent quantities only relevant in the $\bm{r}$ scale and the stratification and Green function only relevant in the $\bm{R}$ scale. The scale separation approximation is not realistic in the subsurface layers (in particular, the mode wavelength is comparable to the typical correlation length associated with turbulence); however, for want of a better alternative, we are led to use this approximation in the following.

Therefore, we make the assumption that the second-order correlation product of the turbulent velocity vanishes for lengths much shorter than the scale associated to the variations of the equilibrium structure. Being able to separate the two scales, as well as the fact that, for radial modes, $X_\omega$ only depends on the radial coordinate, allows us to rewrite the leading term as

\begin{multline}
\left\langle \left|\widehat{\vv_\text{osc}}(\omega)\right|^2\right\rangle = \dfrac{\omega^2}{r_\text{o}^2 c_\text{o}^2 \rho_0(r_\text{o})} \\
\times\displaystyle\int \diff m~\left|\dfrac{\diff X_\omega}{\diff R}\right|^2 \rho_0(R) \displaystyle\int \diff^3\bm{r}~\left\langle \widehat{u_r^2}(\bm{0},\omega)\widehat{u_r^2}^{\star}(\bm{r},\omega)\right\rangle~,
\end{multline}
where we have dropped the variable $\bm{R}$ in favor of the more practical mass variable $m$ . We note that we can only perform this change of variable because the wave equation is radial so that the function $X_\omega(\bm{r})$ only depends on the radial coordinate.

In the following, we focus on establishing the expression of the integral over the fast variable $\bm{r}$. By definition of the temporal Fourier transform appearing in said integral, we have

\begin{multline}
\displaystyle\int \diff^3\bm{r}~\left\langle \widehat{u_r^2}(\bm{0},\omega)~\widehat{u_r^2}^{\star}(\bm{r},\omega)\right\rangle \\
= \dfrac{1}{(2\pi)^2}\displaystyle\iint \diff^3\bm{r}\diff\tau~e^{-j\omega\tau}\left\langle u_r^2(\bm{0},0)~u_r^2(\bm{r},\tau)\right\rangle~.
\end{multline}

We then use the Quasi-Normal Approximation (hereby abbreviated QNA), under which any fourth-order correlation product can be decomposed into a sum of three second-order correlation products, so that \citep{book_lesieur}

\begin{equation}
\left\langle u_r^2(\bm{0},0)u_r^2(\bm{r},\tau)\right\rangle = 2\left\langle u_r(\bm{0},0)u_r(\bm{r},\tau)\right\rangle^2 + \left\langle u_r(\bm{0},0)\right\rangle^2 \left\langle u_r(\bm{r},\tau)\right\rangle^2.
\end{equation}

The last term does not depend on $\tau$ or $\bm{r}$ if the turbulence is homogeneous and uniform, and thus yields zero when the Fourier transform is performed. We can then write

\begin{multline}
\displaystyle\int \diff^3\bm{r}~\left\langle \widehat{u_r^2}(\bm{0},\omega)\widehat{u_r^2}^{\star}(\bm{r},\omega)\right\rangle \\
= \dfrac{2}{(2\pi)^2}\displaystyle\iint \diff^3\bm{r} \diff\tau~e^{-j\omega\tau} \left\langle u_r(\bm{0},0)u_r(\bm{r},\tau)\right\rangle^2 .
\end{multline}

Using the Parseval identity, we can express this as an integral over wave vectors $\bm{k}$ and angular frequencies $\omega$

\begin{multline}
\displaystyle\int \diff^3\bm{r}~\left\langle \widehat{u_r^2}(\bm{0},\omega)\widehat{u_r^2}^{\star}(\bm{r},\omega)\right\rangle = 2\times(2\pi)^2 \\
\times \displaystyle\iint \diff^3\bm{k}\diff\omega'~\text{TF}\left[e^{-j\omega\tau} \left\langle u_r(\bm{0},0)u_r(\bm{r},\tau)\right\rangle\right]\text{TF}\left[\left\langle u_r(\bm{0},0)u_r(\bm{r},\tau)\right\rangle\right]~,
\end{multline}
where the notation $\text{TF}[.]$ refers to temporal and spatial Fourier transform.

We then proceed to describe the second-order correlation product not in terms of time and space increments, but in terms of angular frequencies $\omega$ and spatial modes $\bm{k}$. We denote the temporal and spatial Fourier transform of the second-order correlation product of the $i$-th and $j$-th component of the turbulent velocity as $\phi_{ij}(\bm{k},\omega)$, so that

\begin{multline}
\displaystyle\int \diff^3\bm{r}~\left\langle \widehat{u_r^2}(\bm{0},\omega)\widehat{u_r^2}^{\star}(\bm{r},\omega)\right\rangle \\
= 8\pi^2 \displaystyle\iint \diff^3\bm{k}\diff\omega'~\phi_{rr}\left(\bm{k},\omega'-\omega\right) \phi_{rr}\left(\bm{k},\omega'\right)~.
\label{eq:integral}
\end{multline}

For isotropic turbulence, $\phi_{ij}$ can be expressed analytically \citep{batchelor_book} as

\begin{equation}
\phi_{ij} = \dfrac{E(k,\omega)}{4\pi k^2}\left(\delta_{ij} - \dfrac{k_i k_j}{k^2}\right)~,
\end{equation}
where $E(k,\omega)$ is the specific turbulent kinetic energy spectrum, $k$, $k_i$ and $k_j$ are the norm, $i$-th component and $j$-th component of the wave vector $\bm{k}$, and $\delta_{ij}$ is the Kronecker symbol. The integration over the solid angle of $\bm{k}$ is straightforward and only an integral over its norm remains. However, solar turbulence close to the photosphere is known to be highly anisotropic. To take this anisotropy into account, we follow the formalism developed by \citet{gough77}. In this formalism, the integral over the solid angle of $\bm{k}$ is simply readjusted by adding an anisotropy factor $G$, given by \citep[see Appendix B in][]{samadiG01}

\begin{equation}
G = \displaystyle\int_{-1}^1 \diff\mu~\left(1-\dfrac{Q^2\mu^2}{(Q^2-1)\mu^2 + 1}\right)^2~,
\label{eq:anisotropy}
\end{equation}
where
\begin{equation}
Q^2 = \dfrac{\left\langle u_x^2 \right\rangle}{\left\langle u_r^2 \right\rangle} = \dfrac{\left\langle u_y^2 \right\rangle}{\left\langle u_r^2 \right\rangle}~,
\end{equation}
$u_x$ and $u_y$ referring to the two horizontal components of the turbulent velocity.

This anisotropy factor depends on the ratio between horizontal and vertical turbulent velocities, and therefore depends on the slow $\bm{R}$ variable - or equivalently, on the mass variable $m$. Under this formalism, the integral over $k$ and $\omega$ remains the same as in the isotropic case.

Following \citet{stein67}, we decompose $E(k,\omega)$ into a spatial part $E(k)$, which describes how the turbulent kinetic energy is distributed among modes of different wave numbers, and a temporal part $\chi_k(\omega)$, which describes the statistical life-time distribution of eddies of wavenumber $k$

\begin{equation}
E(k,\omega) = E(k)\chi_k(\omega)~.
\end{equation}

Finally, the integral given by Eq. \eqref{eq:integral} can be rewritten as

\begin{multline}
\displaystyle\int \diff^3\bm{r}~\left\langle \widehat{u_r^2}(\bm{0},\omega)\widehat{u_r^2}^{\star}(\bm{r},\omega)\right\rangle \\
= 2\pi G\displaystyle\int \diff k~\dfrac{E(k)^2}{k^2}\displaystyle\int \diff\omega'~\chi_k(\omega'-\omega)\chi_k(\omega')~.
\label{eq:must_convol}
\end{multline}

As mentioned in the main body of the paper, we have followed two different leads to model the functions $E(k)$ and $\chi_k(\omega)$ in this study. In the following, we only detail what we refer to as the `theoretical spectrum' model, which is based on theoretical prescriptions.

Based on the assumption that turbulent flows are self-similar, Kolmogorov's theory of turbulence leads to a spatial spectrum $E(k) \propto k^{-5/3}$ in the inertial range, between $k=k_0$ (where $k_0$ is the scale at which the kinetic energy is injected in the turbulent cascade, and is henceforth referred to as the injection scale) and the dissipation scale (at which the turbulent kinetic energy is converted into heat). Given the very high Reynolds number characterising solar turbulence ($\mathrm{Re} \sim 10^{14}$), we cast the dissipation scale to infinity. Then, following \citet{musielak94}, we extend the turbulent spectrum below the injection scale by considering that $E(k)$ takes a constant value for $k < k_0$. This extended spectrum, referred to as the broadened Kolmogorov Spectrum (BKS thereafter) was introduced to account for the broadness of the maximum of $E(k)$. Finally, the BKS can be written thus:

\begin{equation}
E(k) = \left\{
\begin{array}{ll}
0.652\dfrac{u_0^2}{k_0} & \text{ if  } 0.2~k_0 < k < k_0 \\
0.652\dfrac{u_0^2}{k_0}\left(\dfrac{k}{k_0}\right)^{-5/3} & \text{ if  } k_0 < k~,
\end{array}
\right.
\label{eq:BKS}
\end{equation}
where $u_0^2 \equiv \left\langle \bm{u}^2(\bm{r}) \right\rangle / 3$ and the $0.652$ factor is introduced so that the total specific kinetic energy of the turbulent spectrum matches $u_0^2 / 2$.

Following \citet{samadi03}, we consider a Lorentzian shape for the temporal spectrum $\chi_k(\omega)$, which is supported both by numerical simulations \citep{samadi03} and by theoretical arguments (if the noise is characterised by a time-correlation function which decays exponentially, its spectrum is Lorentzian). Thus:

\begin{equation}
\chi_k(\omega) = \dfrac{1}{\pi\omega_k}\dfrac{1}{1 + (\omega/\omega_k)^2}~.
\end{equation}

The width of the Lorentzian is the inverse of the typical correlation time-scale and by dimensional arguments, it is proportional to $ku_k$, where $u_k$ is the typical velocity of eddies of wavenumber $k$. However, there remains a substantial indetermination on the actual value of $\omega_k$. To account for this indetermination, we follow \citet{balmforth92} and introduce the dimensionless parameter $\lambda$, so that

\begin{equation}
\omega_k = 2ku_k / \lambda~.
\end{equation}

For a Kolmogorov spectrum, $u_k$ scales as $k^{-1/3}$, which means that we have

\begin{equation}
\omega_k = \omega_{k_0}\left(\dfrac{k}{k_0}\right)^{2/3} \equiv \dfrac{2k_0 u_{k_0}}{\lambda}\left(\dfrac{k}{k_0}\right)^{2/3}~,
\end{equation}

There is a temptation  to approximate the typical velocities of eddies of wavenumber $k_0$ with $u_0$. This assumption requires some discussion, however. Indeed, \citet{stein67} has pointed out that eddies of all sizes have the same Eulerian velocity fluctuations $u_0$. As far as Lagrangian fluctuations go, the fluctuations $u_k$ can be expressed as \citep{stein67}

\begin{equation}
u_k^2 = \displaystyle\int_k^{2k} \diff k~E(k)~.
\end{equation}

Using the expression of $E(k)$ given in Eq. \eqref{eq:BKS} and applying it to $k = k_0$, we finally find $u_{k_0} = 0.602u_0$. Under all these assumptions, all further calculations being carried out, we ultimately obtain the leading term,

\begin{multline}
\left\langle \left|\widehat{\vv_\text{osc}}(\omega)\right|^2\right\rangle = \dfrac{0.353 \lambda\omega^2}{r_\text{o}^2 c_\text{o}^2 \rho_0(r_\text{o})}\displaystyle\int \diff m~\left[\rho_0 G \left|\dfrac{\diff X_\omega}{\diff r}\right|^2 \dfrac{u_0^3}{k_0^4}\right.\\
\left.\left(\displaystyle\int_{0.2}^1~f_1(K)\diff K + \displaystyle\int_1^\infty~f_2(K)\diff K\right)\right]~,
\label{eq:dominant}
\end{multline}
with
\begin{equation}
\begin{array}{l}
f_1(K) = \dfrac{K^{-8/3}}{1 + \left(\dfrac{\lambda\omega}{2.408 u_0 k_0}\right)^2 K^{-4/3}} \\
f_2(K) = \dfrac{K^{-6}}{1 + \left(\dfrac{\lambda\omega}{2.408 u_0 k_0}\right)^2 K^{-4/3}}~,
\end{array}
\end{equation}
and where $K$ is the reduced inverse eddy scale ($K \equiv k/k_0$). We note that all the terms appearing in the integrand depend on the mass variable $m$, particularly $u_0$ and $k_0$, even when the dependence does not appear explicitly. Free parameters are left in this description of the leading term in the form of $\lambda$ and $k_0(m)$; solar turbulence close to the photosphere being as poorly constrained as it is today, we cannot hope to achieve a non-parametrised description of stochastically excited modes of oscillation.

\subsection{The cross term}

Similarly to the leading term, the cross term in the development of $P(\omega)$ shown in Eq. \eqref{eq:dev} can be written as

\begin{multline}
\text{Re}\left(\displaystyle\int \diff\Omega~\widetilde{h}(\mu) \left\langle \widehat{\vv_\text{osc}}(\omega)\widehat{u_n}^{\star}(\omega) \right\rangle\right) = -\dfrac{\omega}{r_\text{o} c_\text{o} \sqrt{\rho_0(r_\text{o})}} \times \\
\displaystyle\int \diff\Omega~\widetilde{h}(\mu)\text{Re}\left(j\displaystyle\int \diff^3\bm{r_\text{s}}~\rho_0\dfrac{\diff X_\omega}{\diff r}\left\langle \widehat{u_r^2(\bm{r_\text{s}})}\widehat{u_n (\bm{r_\text{o}})}^\star \right\rangle\right)~,
\label{eq:correl}
\end{multline}
where $\mu$ is the cosine of the angle between the local vertical direction and the direction of the line of sight.

We note that this time, one of the velocities in the correlation product is estimated at a fixed location corresponding to the observation height, so that only one variable is left. Expressing the line-of-sight component of $\bm{u}$ as $u_n = u_r\cos\theta - u_\theta \sin\theta$, this transforms into

\begin{multline}
\text{Re}\left(\displaystyle\int \diff\Omega~\widetilde{h}(\mu) \left\langle \widehat{\vv_\text{osc}}(\omega)\widehat{u_n}^{\star}(\omega) \right\rangle\right) = -\dfrac{\omega}{r_\text{o} c_\text{o} \sqrt{\rho_0(r_\text{o})}} \times \\
\left[ \text{Re}\left(j\displaystyle\int \diff m~\dfrac{\diff X_\omega}{\diff r}\left\langle \widehat{u_r^2(\bm{r_\text{s}})}\widehat{u_r (\bm{r_\text{o}})}^\star \right\rangle\right) \displaystyle\int \diff\Omega~\widetilde{h}(\mu)\mu \right. \\
\left. + \text{Re}\left(j\displaystyle\int \diff m~\dfrac{\diff X_\omega}{\diff r}\left\langle \widehat{u_r^2(\bm{r_\text{s}})}\widehat{u_\theta (\bm{r_\text{o}})}^\star \right\rangle\right) \displaystyle\int \diff\Omega~\widetilde{h}(\mu)\sqrt{1-\mu^2} \right]~.
\end{multline}

Since $u_\theta$ is a horizontal component of the turbulent velocity, if we consider there is no preferential horizontal direction as far as turbulence goes, the third-order correlation product appearing in the second integral cancels out, so that we are left with the first integral only. The latter can be rewritten thus:

\begin{multline}
\displaystyle\int \diff m~\dfrac{\diff X}{\diff r}\left\langle \widehat{u_r^2(\bm{r_\text{s}})}\widehat{u_r (\bm{r_\text{o}})}^\star \right\rangle \\
= \displaystyle\int \diff^3\bm{r_{\text{s}1}} \diff^3\bm{r_{\text{s}2}}~\rho_0(\bm{r_{\text{s}1}}) \dfrac{\diff X}{\diff r}(\bm{r_{\text{s}1}}) \left\langle \widehat{u_r^2(\bm{r_{\text{s}1}})}\widehat{u_r (\bm{r_{\text{s}2}})}^\star \right\rangle \delta(\bm{r_{\text{s}2}}-\bm{r_\text{o}})~,
\end{multline}
where we have artificially introduced a second spatial variable $\bm{r_{\text{s}2}}$, so as to get an expression formally similar to that of the leading term above. Performing the same change of variables, and plugging the definition of the temporal Fourier transform, we have

\begin{multline}
\displaystyle\int \diff m~\dfrac{\diff X}{\diff r}\left\langle \widehat{u_r^2(\bm{r_\text{s}})}\widehat{u_r (\bm{r_\text{o}})}^\star \right\rangle \\
= \dfrac{1}{(2\pi)^2}\rho_0(\bm{r_\text{o}}) \dfrac{\diff X}{\diff r}(\bm{r_\text{o}}) \displaystyle\int \diff\tau \diff^3\bm{r}~e^{-j\omega\tau} \left\langle u_r^2(0,\bm{0})u_r(\tau,\bm{r}) \right\rangle~.
\label{eq:before_PCM}
\end{multline}

The challenge in estimating the contribution of correlated turbulent perturbations fundamentally lies in the correct determination of non-local, third-order correlation products. While the QNA provides an adequate closure relationship for fourth-order moments, it yields vanishing third-order moments, which is known to lead to serious violations of the energy conservation principle, as well as an impossibility for the turbulent cascade to develop \citep[see][]{kraichnan57}.

In order to estimate the third-order moments, we follow \citet{belkacem06} and use the Plume closure model, which consists of separating the flow into upward flows and downward plumes, each normally distributed, with different mean values and standard deviations. In addition, we consider that the downflows are much more turbulent than the upflows \citep[which is supported by][according to whom the intergranular lanes harbour stronger turbulence than the granules themselves at the Sun's surface]{goode98}, and that the two types of flows considered separately have zero third-order correlation products. In \citet{belkacem06}, the authors use the same approximations but focused on one-point correlation products; however, the calculations can be easily extended to the two-point correlation products that we need and the model yields

\begin{multline}
\langle u_r(\bm{R},t)^2 u_r(\bm{R}+\bm{r},t+\tau)\rangle = \left[a(1-a)^3 - a^3(1-a)\right] \delta u^3 \\
-a(1-a)\left[2\langle \widetilde{u_d}(\bm{R},t) \widetilde{u_d}(\bm{R}+\bm{r},t+\tau) \rangle + \langle \widetilde{u_d}(\bm{R},t)^2 \rangle\right]\delta u~,
\label{eq:PCM_app}
\end{multline}
where $a$ is the relative horizontal section of the upflows, $\delta u$ is the difference between the mean velocity of the two types of flows (considering their respective signs, it actually is the sum of their absolute values), and $\widetilde{u_d}$ is the fluctuation of the downflow velocity around its mean value. The only additional approximation we make to adapt these calculations to two-point correlation products is that the parameters of the model $a$ and $\delta u$ vary on scales much greater than the typical correlation length, which is another illustration of the scale separation approximation, which we have already used for the leading term (see above).

Note that strictly speaking, the third-order moment appearing in Eq. \eqref{eq:PCM_app} and yielded by the PCM are centred. However, we consider that the mean value of the overall vertical velocity of the flow is sufficiently low (compared to its standard deviation for instance) to be neglected. Therefore, the moment described by Eq. \eqref{eq:PCM_app} may interchangeably refer either to a centred or non-centred moment.

Also note that this closure relation is written here in terms of $\widetilde{u_d}$ (i.e. the turbulent fluctuations in the downflows only). It would be more practical to rewrite it in terms of $u_r$ (i.e. the total turbulent fluctuations). The two are related through
\begin{equation}
\langle \widetilde{u_d}(\bm{R}, t)\widetilde{u_d}(\bm{R}+\bm{r}, t+\tau) \rangle = \dfrac{1}{1-a} \langle u_r(\bm{R}, t)u_r(\bm{R}+\bm{r}, t+\tau) \rangle - a\delta u^2~.
\label{eq:PCM_conv}
\end{equation}

Plugging Eq. \eqref{eq:PCM_conv} into Eq. \eqref{eq:PCM_app}, we obtain a closure model that allows us to write the third-order moments as a function of second-order moments only, after which we can use the same prescriptions for turbulence as we did for the leading term. We note that while this closure relation contains many terms, not many survive the Fourier transform in Eq. \eqref{eq:before_PCM} as all terms not depending on $\bm{r}$ or $\tau$ will not contribute to the Fourier transform. The integral over $m$ appearing in \eqref{eq:correl} becomes

\begin{equation}
\displaystyle\int \diff m~\dfrac{\diff X}{\diff r}\left\langle \widehat{u_r^2(\bm{r_s})}\widehat{u_r (\bm{r_o})}^\star \right\rangle = -8\pi^2 a\delta u \rho_0(r_o) \dfrac{\diff X}{\diff r}(r_o)\phi_{rr}(\bm{0},\omega)~.
\end{equation}

Formally, the integration over $\bm{r}$ makes the value of $\phi_{rr}$ at $\bm{k}=\bm{0}$ appear. Physically, that means only the largest eddies have a significant impact on the correlation between the mode and the turbulent perturbations. Since considering eddies characterised by $\bm{k} = \bm{0}$  does not actually make physical sense, we considered a comprise  by assuming that the largest contributing eddies are those at the injection scale $k_0$, so that the correlated turbulent perturbations term finally becomes

\begin{multline}
\text{Re}\left(\displaystyle\int \diff\Omega~\widetilde{h}(\mu) \left\langle \widehat{\vv_\text{osc}}(\omega)\widehat{u_n}^{\star}(\omega) \right\rangle\right) = C(\omega)\displaystyle\int \diff\Omega~\widetilde{h}(\mu)\mu \\
C(\omega) = -\dfrac{1.083 \lambda a\omega u_0 \delta u}{r_\text{o} c_\text{o} k_0^4} \\
\times \sqrt{\rho_0(r_\text{o})}\left.\dfrac{\diff\text{Im}X_\omega}{\diff r}\right|_{r_o} \times\left(1 + \left(\dfrac{\lambda\omega}{1.204 k_0 u_0}\right)^2\right)^{-1}~,
\label{eq:crossed}
\end{multline}
where every parameter is estimated at the observation height $r_\text{o}$, even when not specified explicitly. Note that we introduce no new free parameter compared to those used for the leading term, as the parameters $a$ and $\delta u$ appearing in the PCM are extracted from numerical hydrodynamic simulations of the stellar atmosphere. Plugging Eqs. \eqref{eq:dominant} and \eqref{eq:crossed} in the expression of the velocity power spectral density given by Eq. \eqref{eq:dev}, we obtain Eq. \eqref{eq:power}.

\section{A simplified toy model}\label{app:toy_model}

Here we consider a simplified model of solar acoustic mode excitation, which has already been developed, used, and analysed in previous works \citep[see e.g.][]{abramsK96, chaplinA99}. In the scope of this toy-model, the acoustic potential appearing in Eq. \eqref{eq:wave_equation} takes the form of a square well, and the sound speed $c$ is considered constant throughout the entire stellar radius. The latter approximation allows us to substitute the radial coordinate $r$ in the wave equation for the acoustic depth $\tau$, defined such that $\diff\tau = \diff r/c$. In this approximation, the wave equation simply yields
\begin{equation}
\dfrac{\diff^2\Psi}{\diff\tau^2} + (\omega^2 - V(\tau) + j\omega\gamma)\Psi = \delta(\tau-\tau_\text{s})~,
\label{eq:waveeq_toy}
\end{equation}
where we have introduced a point-like source at acoustic depth $\tau_\text{s}$, and the acoustic potential is
\begin{equation}
V(\tau) = \left\{
\begin{array}{ll}
+\infty &\text{ if } \tau < 0 \\
0 &\text{ if } 0 \leqslant \tau \leqslant a \\
\alpha^2 &\text{ if } a < \tau~.
\end{array}
\right.
\end{equation}

In this model, $a$ represents the acoustic length of the cavity (for radial modes, it corresponds to the time it takes for sound waves to propagate throughout the entire stellar radius) and $\alpha$ is the acoustic cut-off angular frequency above which waves are no longer confined. We added an infinite step at $\tau = 0$ to force the wave variable $\Psi$ to vanish at the centre.

\subsection{Analytic solution of Eq. \eqref{eq:waveeq_toy}}

In order to solve the wave equation, it should be rewritten in terms of matrices:

\begin{equation}
\dfrac{\diff X}{\diff\tau} = AX + B~,
\label{eq:matrix_eq}
\end{equation}
where

\begin{equation}
\begin{array}{l}
X = \begin{bmatrix} \Psi \\ \diff\Psi / \diff\tau \end{bmatrix} \\
\vspace{1pt} \\
A = \begin{bmatrix} 0 & 1 \\ V(\tau) - \omega^2 - j\omega\gamma & 0 \end{bmatrix} \\
\vspace{1pt} \\
B = \begin{bmatrix} 0 \\ \delta(\tau-\tau_\text{s}) \end{bmatrix}~.
\end{array}
\end{equation}

The general solution to the homogeneous equation is $X_h(\tau) = \exp(A\tau) C$, where $C$ is a constant vector. A particular solution to the inhomogeneous equation can then be sought in the form $X_p(\tau) = \exp(A\tau) C(\tau)$. For each domain in which the matrix $A$ is constant, injecting this form in Eq. \eqref{eq:matrix_eq} yields

\begin{equation}
C(\tau) = \left\{
\begin{array}{ll}
\begin{bmatrix} 0 \\ 0 \end{bmatrix} & \text{ if } \tau < \tau_\text{s} \\
\vspace{1pt} \\
\begin{bmatrix} 0 \\ \exp(-A\tau_\text{s}) \end{bmatrix} & \text{ if } \tau_\text{s} \leqslant \tau~.
\end{array}
\right.
\end{equation}

$A$ being piecewise constant, we can thus write the general solution to Eq.\eqref{eq:matrix_eq} as

\begin{equation}
X(\tau) = \left\{
\begin{array}{ll}
\exp(A\tau) C & \text{ if } \tau < \tau_\text{s} \\
\exp(A\tau) \left(C + \begin{bmatrix} 0 \\ \exp(-A\tau_\text{s}) \end{bmatrix} \right) & \text{ if } \tau_\text{s} \leqslant \tau~,
\end{array}
\right.
\label{eq:general_sol}
\end{equation}
where the integration constant $C$ is constant on whichever domain $A$ is; in other words, $C$ is piecewise constant on the same domains as $A$, that is, between $0$ and $a$, and above $a$ separately. In the following, we denote the column vector $C$ as $\left[ A_\text{i} ~ B_\text{i} \right]$ in the former domain, and $\left[ A_\text{o} ~ B_\text{o} \right]$ in the latter, the indices i and o referring to the inner and outer regions, respectively.

The simple form of $A$ allows for a straightforward computation of its exponential. We obtain

\begin{equation}
\exp(A\tau) = \left\{
\begin{array}{ll}
\begin{bmatrix} \cos \omega_\text{i}\tau & \dfrac{\sin \omega_\text{i} \tau}{\omega_\text{i}} \\ -\omega_\text{i} \sin \omega_\text{i} \tau & \cos \omega_\text{i}\tau \end{bmatrix} & \text{ if } 0 < \tau < a \\
\vspace{1pt} \\
\begin{bmatrix} \cosh \omega_\text{o}\tau & \dfrac{\sinh \omega_\text{o} \tau}{\omega_\text{o}} \\ \omega_\text{o} \sinh \omega_\text{o} \tau & \cosh \omega_\text{o}\tau \end{bmatrix} & \text{ if } a < \tau~,\\
\end{array}
\right.
\end{equation}
where $\omega_\text{i}^2 = \omega^2 + j\omega\gamma$ and $\omega_\text{o}^2 = \alpha^2 - \omega^2 - j\omega\gamma$ (with the understanding that $0 < \omega < \alpha$).

Finally, injecting this into the general solution \eqref{eq:general_sol} and only keeping the first line (at this point, the second one only gives the derivative of the solution and is redundant with the solution itself) yields the following expression, depending on whether the source is inside the cavity or not:

\begin{equation}
\Psi_\text{i}(\tau) = \left\{
\begin{array}{ll}
A_\text{i}\cos \omega_\text{i}\tau + \dfrac{B_\text{i}}{\omega_\text{i}}\sin \omega_\text{i}\tau & \text{ if } \tau < \tau_\text{s} \\
\begin{split}
A_\text{i}\cos \omega_\text{i}\tau + \dfrac{B_\text{i}}{\omega_\text{i}}\sin \omega_\text{i}\tau \\
+ \dfrac{1}{\omega_\text{i}}\sin \omega_\text{i}(\tau-\tau_\text{s}) 
\end{split} & \text{ if } \tau_\text{s} < \tau < a \\
A_\text{o} \cosh \omega_\text{o}\tau + \dfrac{B_\text{o}}{\omega_\text{o}}\sinh \omega_\text{o}\tau & \text{ if } a < \tau_\text{s}~,
\end{array}
\right.
\label{eq:sol_i}
\end{equation}
and

\begin{equation}
\Psi_\text{o}(\tau) = \left\{
\begin{array}{ll}
A_\text{i}\cos \omega_\text{i}\tau + \dfrac{B_\text{i}}{\omega_\text{i}}\sin \omega_\text{i}\tau & \text{ if } \tau < a \\
A_\text{o} \cosh \omega_\text{o}\tau + \dfrac{B_\text{o}}{\omega_\text{o}}\sinh \omega_\text{o}\tau & \text{ if } a < \tau < \tau_\text{s} \\
\begin{split}
A_\text{o} \cosh \omega_\text{o}\tau + \dfrac{B_\text{o}}{\omega_\text{o}}\sinh \omega_\text{o}\tau \\
+ \dfrac{1}{\omega_\text{o}}\sinh \omega_\text{o}(\tau-\tau_\text{s})
\end{split} & \text{ if } \tau_\text{s} < \tau~,\\
\end{array}
\right.
\label{eq:sol_o}
\end{equation}
where $\Psi_\text{i}$ is the solution if the source is inside the cavity ($\tau_\text{s} < a$) and $\Psi_\text{o}$ is the solution if the source is outside ($a < \tau_\text{s}$). Predictably, this general solution contains 4 constants of integration (in the form of $A_\text{i}$, $B_\text{i}$, $A_\text{o}$ and $B_\text{o}$), and we therefore need 4 boundary conditions to find $\Psi_\text{i}$ and $\Psi_\text{o}$. We impose that $\Psi(\tau=0) = 0$ and that the solution do not diverge when $\tau \rightarrow +\infty$; furthermore, we impose that both $\Psi$ and its derivative be continuous at $\tau = a$. With all calculations having been carried out, this set of boundary conditions finally gives us:

\begin{equation}
\begin{array}{l}
\Psi_{\text{i,o}}(\tau) = -\dfrac{f_{\text{i,o}}(\tau_\text{s})}{\omega_\text{i}\cos\omega_\text{i} a + \omega_\text{o}\sin \omega_\text{i} a}\exp^{-\omega_\text{o}(\tau - a)} \\
\vspace{1pt} \\
f_\text{i}(\tau_\text{s}) = \sin \omega_\text{i}\tau_\text{s} \\
\vspace{1pt} \\
\begin{split}
f_\text{o}(\tau_\text{s}) = \omega_\text{i} \cos \omega_\text{i} a \sinh\omega_\text{o}(\tau_\text{s}-a) \\
+ \omega_\text{o} \sin \omega_\text{i} a \cosh \omega_\text{o}(\tau_\text{s}-a)~.
\end{split}
\end{array}
\end{equation}

We note that the above expression is only valid if $\tau > a$ and $\tau > \tau_\text{s}$. Since the first condition means that we observe the mode of oscillation above the upper turning point (which is always the case in practice) and since the second condition means that the excitation of the mode occurs in layers located deeper than the observation height in the atmosphere (which is also always the case in practice, at least for spectrometric measurements), these are not restrictive conditions.

\subsection{Discussion}

In each case (source inside or outside the cavity), the solution $\Psi_{\text{i,o}}$ can be decomposed into three parts:
\begin{itemize}
\item the denominator corresponds to the Wronskian $W(\omega)$ of the wave equation. $|1 / W(\omega)|^2$ peaks at the eigenfrequencies associated to the acoustic cavity and is responsible for the presence of resonant modes in the spectrum. The line profile it generates have a Lorentzian profile, so long as the damping rate of the modes are much smaller than their frequency;
\item the exponential part accounts for the fact that the modes are evanescent outside the acoustic cavity, so that the higher in the atmosphere the mode is observed, the lower its amplitude as we observe it. Therefore, it only affects the observed amplitude of the mode, not its line profile;
\item the numerator $f_{\text{o,i}}(\tau_\text{s})$ is responsible for the mode line profile asymmetry. Because it is the only term that depends on the source position $\tau_\text{s}$, it is commonly said that mode asymmetry is caused by source localisation.
\end{itemize}

Regardless of whether the source is located inside or outside the cavity, it can be seen from Eqs. \eqref{eq:sol_i} and \eqref{eq:sol_o} that the third term $f_{\text{o,i}}(\tau_\text{s})$ actually corresponds to the amplitude of the solution $\Psi_{\text{i,o}}$ at $\tau = \tau_\text{s}$. This leads us to the following conclusion, which we reproduce in the main body of the paper: for a given frequency, the amplitude of the excited wave is proportional to the radial profile associated with the wave at the source of excitation.

\end{appendix}

\end{document}